\documentclass[twoside,11pt]{article}
\usepackage{amsmath,amssymb,amsfonts}
\usepackage{amsfonts}
\usepackage{graphicx}
\usepackage{graphics}
\usepackage{breqn}
\usepackage{microtype}
\usepackage{setspace}
\begin{document}

\setlength{\unitlength}{10mm}
\newcommand{\f}{\frac}
\newtheorem{theorem}{Theorem}[section]
\newcommand{\sta}{\stackrel}

\title{Lagrange multiplier and Wess-Zumino variable as extra dimensions in the torus universe}

\author{
Salman Abarghouei Nejad
\\
{\scriptsize Department of Physics,
}\\{\scriptsize University of Kashan, Kashan 87317-51167, I. R. Iran}
\\
\\
Mehdi Dehghani
\\
{\scriptsize Department of Physics, Faculty of Science,
}\\{\scriptsize Shahrekord University, Shahrekord, P. O. Box 115, I. R. Iran}
\\
\\
Majid Monemzadeh
\\
{\scriptsize Department of Physics,
}\\{\scriptsize University of Kashan, Kashan 87317-51167, I. R. Iran}}

\date{}
\maketitle
\vspace{1cm}
\vspace{-1.5cm}\begin{abstract}
\noindent  We study the effect of the the simplest geometry which is imposed via the topology of the universe by gauging non-relativistic particle model on torus and 3-torus with the help of symplectic formalism of constrained systems. Also, we obtain generators of  gauge transformations for gauged models. Extracting corresponding Poisson structure of  existed constraints, we show the effect of the shape of the universe on canonical structure of phase-spaces of models and suggest some phenomenology to prove the topology of the universe and probable non-commutative structure of the space. In addition, we show that the number of extra dimensions in the phase-spaces of gauged embedded models are exactly two. Moreover, in classical form, we talk over modification of Newton's second law in order to study the origin of the terms appeared in the gauged theory.



\end{abstract}

\newpage
\section{Introduction}

\subsection{Why torus?}

The torus universe model or the doughnut theory of the universe, is the model which describes the universe as a doughnut, having surface with topology \footnote{The word topology in this article is used as the global shape and characteristics of the universe and we do not intended its pure mathematical definition, where properties of space that are preserved under continuous deformations are studies.} of a three dimensional torus. Historically, the first explanations of the shape of the universe were proposed in the mid 60s, after the discovery of CMB by Starobinsky and Zeldovich \cite{PC-12-61}.

In experimental point of view, data of cosmos radiation measurements gathered by satellite COBE, shows small discrepancies in temperature fluctuation. This shows that the universe consists of regions of varying densities. Stenemse and Silk proposed that this paradox, i.e. the isotropic universe with different regional densities, suggests that universe may have a complicated geometric structure \cite{TA3}. In other words, these fluctuations show that multiply connected universes are possible, and the simplest one is a 3-torus \cite{TA3,Cornish0}. Also, simulations of CMB map and the angular power spectrum of temperature fluctuations, considering the torus topology, and comparing them with the observations of the COBE satellite in order to obtain the lower limit of universe size, suggest that we live in a small universe \cite{Starobinsky,Uzan1,Uzan2,Riazuelo,Aurich}.

On the other hand, although there is no statistically significant evidence to support what the topology of the universe would be, there are some suggestions which talk over a 3-torus as the probable shape of the universe \cite{Cornish1,Cornish2,Cornish3,Cornish4,Aurich3}.

Moreover, data gathered by WMAP satellite shows more intense CMB across one plane of the universe in comparison with others, which forms a straight line in the universe. Where radiation surpasses its quota for the size of the plane seen, one can say that the universe has overflowed in that direction and creates a plane in other directions. Thereby, the invisible loop of a torus may have been created perpendicular to the direction of the plane. Thus, the analysed CMB maps from data obtained from WMAP has released some results in favour of a torus form of the universe \cite{PC-12-61,TA5,Aurich2}. Measurements of WMAP shows that the universe is flat with only $0.4 \%$ margins of error. On the other hand, flat universes with boundaries or edges are not desired mathematically, and thus, they are excluded from consideration. Although there are some finite compact universe models without boundaries, the torus universe is the one which explains an overall flat and a finite universe \cite{TD9}.

Theoretically, string theory and also theories considering extra large dimensions suggest that we live in a universe with higher dimensions of space-time and most of the modern cosmological models are founded on such assumptions. Moreover, the problems of the standard cosmology are avoided, considering higher dimensional space–time, and also most of the predictions of the inflation cosmology are fulfilled via these approaches \cite{TB1,TB2,TB3}.

In order to combine topological theories and extra large dimensions universe, it has been shown that cyclic universe models can be acquired in a toroidal space–time which is embedded in a five-dimensional bulk with large extra dimensions, and the three dimensional space has been shown as a closed ring, moving on the surface of the torus \cite{TB,TC19}.

If we expect that the universe has topology of a torus, we can construct a gauge theory, using the Lagrangian of a particle on the torus, and quantize such a gauge theory, and extract its gauge transformation relations.
Our goal to study the motion of a non-relativistic particle on a torus and gauging that model is to obtain a configuration space with extra dimensions. As we know, studying the motion of a free particle is the most powerful laboratory, in which we can test whether the torus universe exists or not. Making such gauge theories and studying its Hamiltonian spectrum may help us to understand the real topology of the universe. Moreover, with investigating the final obtained phase-space, one can check the commutativity and non-commutativity of the universe. In addition, we can determine the ratio of two diameters of the torus.

Another point of view in which we can study the constructed classical theory on a torus is the modification of Newton's law, that talks over the corrections added to the Newtonian classical mechanics. In the common Poisson structure, Hamiltonian equations of motion and Newton laws are equivalent \cite{Goldestein}. In this article, we construct a classical theory which has an unusual Poisson structure due to its constrained structure. This Poisson structure adds some additional terms to the Hamiltonian and consequently to the equations corresponding Newton's second law which can be studied via the MOND phenomenological theory.

Our tool to construct a gauge theory which reduces to a particle on the torus after gauge fixing is the symplectic gauge analysis approach will be discussed later.

\subsection{Gauge theories and constraints}

As we know, gauge invariance is one of the most significant and practical concepts in theoretical physics. This concept is the cornerstone of the standard model of elementary particles. Gauge invariance is due to the presence of the important physical variables which are independent of local reference frames \cite{G1}. Whenever a change is applied in an arbitrary reference frame, which makes changes in such variables, gauge transformation  occurs. Such physical variables are called gauge invariant variables.

Generally, we deal with gauge invariance, or in other words, local invariance, which produces gauge bosons in fundamental interactions. As a physical law, the existence of (local) gauge symmetry in particle physics is the sign of the presence of interactions \cite{thooft1}.

It is very important to know that quantization of gauge theories entails a particular prudence, because of the presence of gauge symmetry exist some nonphysical degrees of freedom that must be eliminated before and after the quantization is applied \cite{G2}.

On the other hand, in a gauge theory, the equations of motion are not able to determine the dynamics of the system thoroughly at every moment. Thus, one of the most particular features of a gauge theory is the emergence of arbitrary time dependent functions in general solutions of the equations of motion. The emergence of such time dependent functions is accompanied by the relations between phase-space coordinates, which are called constraints \cite{G1,Bergmann}.

In order to quantize such systems, identities between phase-space coordinates are classified into two main groups by Dirac \cite{Dirac1}.
The first group are identities which are present in phase-space, similar to a coordinate or a momentum variable. These identities, which transform the physical system without any changes in phase-space, are called first-class constraints, and according to Dirac's guess are generators of gauge transformations in phase-space. The second group are not related to any degree of freedom and must be removed. Presence of such identities, which are called second-class constraints, indicates the absence of gauge symmetry in the system. Therefore, to gauge a system, containing second-class constraints, we must transform them to first-class ones, as a first step \cite{Dirac2,Monem1}.

There are some approaches to perform such a conversion, like BFT method \cite{BFFT1, BFFT2, BFFT3,Monem2, BFFT4}, the symplectic formalism \cite{G2, FJ1, FJ2, FJ3}, and the Noether dualization technique \cite{Noether1, Noether2, Noether3}.
As we mentioned before, in order to gauge a system with second-class constraints, we use the symplectic approach in order to embed a non-invariant system in an extended phase-space \cite{G4, G5,Monem4}.

\subsection{Symplectic Formalism}

Symplectic formalism was introduced by Faddeev and Jackiw \cite{FJ1}, to avoid consistency problems which spoil the Poisson brackets algebra and consequently fail any quantization techniques in constrained systems \cite{G7,Monem3}. The mathematics of this formalism is based on the symplectic structure of the phase-space, and therefore, is different from other approaches. Also, in the symplectic formalism there is no distinction between the first and second-class constraints as in the case of the other quantization procedures \cite{FJ3}.

The starting point of the symplectic approach is a Lagrangian which is first order in the time derivatives. All second order Lagrangian terms can be converted to first order ones by enlarging the corresponding configuration space so that it includes the conjugate momentum of the coordinate variables \cite{G9}. Being dependent only on first order Lagrangian makes the symplectic approach independent from the classification of the constraints into primary, secondary, etc. \cite{kim}.
In this approach, instead of solving the constraints, one adds their time derivatives to the Lagrangian and considers corresponding Lagrange multipliers as additional coordinates \cite{Mojiri}. Also, to convert the nature of second-class constraints to first ones, the phase-space would be extended with the help of Wess-Zumino variables \cite{WZ}. After such a conversion, choosing conventional zero-modes which are generators of gauge transformations and obey particular boundary conditions, one can eliminate Wess-Zumino variables, which makes the gauged model equivalent to the original system \cite{Henneaux}.

\section{Gauging a non-relativistic particle model on the torus}

\subsection{Particle on the torus}

In the first part of this article we assume a non-relativistic particle on a torus in a three-dimensional configuration space as a toy model. Considering this model, the particle lives on a two-dimensional configuration space, effectively. After our gauging process, we will see that at least one dimension is added to the previous configuration space, which makes its space more realistic.

In all sections of this article we get the radii of the torus 1 and $ \varsigma $, in order to use dimensionless coordinates.
Thus, in spherical coordinates, the surface of a torus is defined by
\begin{eqnarray}
  \nonumber x &=& (1+\varsigma\cos\theta)\cos\varphi \\
  \nonumber y &=& (1+\varsigma\cos\theta)\sin\varphi \\
  z &=& \varsigma \sin\theta
\end{eqnarray}

The surface of the torus is described by primary constraint $\phi_1=0$ in configuration space for free particle on it.
\begin{equation}\label{phi1}
\phi_1(r,\theta) =r^2-2\varsigma\cos\theta-(1+\varsigma^2),
\end{equation}
where, $ r^2=x^2+y^2+z^2 $. In this coordinate canonical Hamiltonian for unit mass is
\begin{equation}\label{ham}
H_c=\frac{1}{2}(p_r^2+\frac{p^2_{\theta}}{r^2}+\frac{p^2_{\varphi}}{r^2\sin^2\theta}).
\end{equation}

In formal constrained analysis we arrive to secondary (final) constraint in phase-space as
\begin{equation}\label{second}
    \phi_2(r,\theta,p_r,p_{\theta}) =2(rp_r+\frac{\varsigma p_{\theta}\sin\theta}{r^2}).
\end{equation}
The set of constraints form a second-class system with non-constant $\Delta$ matrix as
\begin{equation}\label{delta}
    \Delta_{12}=4(r^2+\frac{\varsigma^2\sin^2\theta}{r^2}),
\end{equation}
which makes its embedding by BFT method problematic. This is the reason that we use the symplectic approach, which is not affected by the Poisson structure of second-class constraints.

\subsection{Symplectic analysis of a particle on the torus}

Constructing first-class models from a singular Lagrangian is more straightforward in the symplectic formalism than other similar approaches. This is done by embedding the primary model in an extended phase-space.

In this model, the singularity nature of the free particle Lagrangian due to its configuration constraint, $ \phi_1(r,\theta)$, can be imposed by a new dynamical variable (say undetermined Lagrange multiplier) $ \lambda$, in such a way that adds the constraints to the free Lagrangian,
\begin{equation}\label{firstLag}
    L^{(0)}=\dot{r}p_r+\dot{\theta}p_{\theta}+\dot{\varphi}p_{\varphi}-H_c-\lambda_{1}\phi_{1}(r,\theta).
\end{equation}

Symplectic variables and their conjugate momenta as the symplectic one-form can be read off from the model straightforwardly \footnote{ In this article the Greek indices, $ \alpha $ , $ \beta $, $ \tilde \alpha $, and $\tilde \beta $, are used to determine phase-space variables.},
\begin{eqnarray}\label{symplectic var 0}
  \nonumber {\xi}^{(0)}_{\alpha} &=& (r,\theta,\varphi,p_r,p_{\theta},p_{\varphi},\lambda_1), \\
  \mathcal A^{(0)}_{\alpha} &=& (p_r,p_{\theta},p_{\varphi},0,0,0,0).
\end{eqnarray}
Symplectic two-form is defined by
\begin{equation}\label{f}
f_{\alpha\beta}=\partial_{\alpha}\mathcal A^{(0)}_{\beta}-\partial_{\beta}\mathcal A^{(0)}_{\alpha}.
\end{equation}
Thus, the zeroth-iterated symplectic two-form, using equation \eqref{f} and symplectic variables and corresponding conjugate momenta \eqref{symplectic var 0}, will be obtained as follows,
\begin{eqnarray}
f^{(0)}_{\alpha\beta}=\begin{pmatrix}
 \textbf{0}_{3\times3} & -\textbf{1}_{3\times3} & \textbf{0}_{3\times1} \\
    \textbf{1}_{3\times3} & \textbf{0}_{3\times3} & \textbf{0}_{3\times1} \\
    \textbf{0}_{1\times3} & \textbf{0}_{1\times3} & 0 \\
\end{pmatrix}.
\end{eqnarray}
This matrix is singular, and so, it has the following null vector,
\begin{eqnarray}
\textit{n}^{(0)}_{\alpha}=\begin{pmatrix}
 \textbf{0}_{1\times3} &  \textbf{0}_{1\times3} & 1
\end{pmatrix} .
\end{eqnarray}

Using the zero iterative potential,
\begin{eqnarray}
\mathcal{V}^{(0)}=H_{c}+\lambda_{1}\phi_{1},
\end{eqnarray}
the first constraint \eqref{phi1} will be obtained from the following formula.
\begin{equation}\label{cons}
\phi_{1}=n_{\alpha}^{(0)}\frac{\partial \mathcal{V}^{(0)}}{\partial \xi^{(0)\alpha}}.
\end{equation}

Substituting the first constraint, obtained from \eqref{cons} into the original Lagrangian, we can put the constraint into the kinetic part of the Lagrangian. It means that we make the primary constraint $ \phi_{1} $ as a momentum conjugate to the variable $ \lambda_{1} $. In other words, we convert the strongly nonlinear constraint, $ \phi_{1} $, into momentum part (linear constraint) of phase-space. Hence, the first iterative Lagrangian will be obtained as
\begin{equation}\label{L1}
L^{(1)}=\dot{r}p_r+\dot{\theta}p_{\theta}+\dot{\varphi}p_{\varphi}-\dot{\lambda}_{1} \phi_{1}-H_{c}.
\end{equation}
We see that the constraint is omitted from the potential. So, for the first iterative potential we have,
\begin{equation}\label{V1}
\mathcal{V}^{(1)}=H_{c}.
\end{equation}

Now, we read off new symplectic variables and one-form from \eqref{L1},
  \begin{eqnarray}
  \nonumber {\xi}^{(1)}_{\alpha} &=& (r,\theta,\varphi,p_r,p_{\theta},p_{\varphi},\lambda_{1}), \\
  \mathcal A^{(1)}_{\alpha} &=& (p_r,p_{\theta},p_{\varphi},0,0,0,\phi_{1}).
\end{eqnarray}
The corresponding symplectic two-form is constructed as,
\begin{eqnarray}\label{f(1)}
f_{\alpha\beta}^{(1)}=
\begin{pmatrix}
   \textbf{0}_{3\times3} & -\textbf{1}_{3\times3} & \textbf{u}_{1\times3}^{T} \\
   \textbf{1}_{3\times3} & \textbf{0}_{3\times3} & \textbf{0}_{3\times 1} \\
  -\textbf{u}_{1\times3} & \textbf{0}_{1\times3} & 0 \\
\end{pmatrix},
\end{eqnarray}
which;
\begin{eqnarray}\label{u}
&& u_{\alpha}=\frac{\partial  \phi_{1}}{\partial q^{\mu}}
 \nonumber \\
&& \hspace{1cm} =\begin{pmatrix}
 2r & 2\varsigma\sin\theta & 0
\end{pmatrix}.
\end{eqnarray}
The two-form \eqref{f(1)} is a singular one and it has following null vectors,
\begin{eqnarray}
\nonumber \textit{n}^{(1)}_{1 \alpha}=\begin{pmatrix}
\textbf{0}_{1\times 3} & \textbf{u}_{1\times3} & 0
\end{pmatrix}, \\
\textit{n}^{(1)}_{2 \alpha}=\begin{pmatrix}
\textbf{0}_{1\times 3} & \textbf{0}_{1\times3} & 1
\end{pmatrix} .
\end{eqnarray}
From linear algebra, we know that the linear combination of these null vectors is also a null vector,
\begin{equation}
\textit{n}_{\alpha}=\textit{n}^{(1)}_{1 \alpha}+h\textit{n}^{(1)}_{2 \alpha}
\end{equation}

Using \eqref{cons}, we obtain the second constraint.
\begin{eqnarray}
\phi_{2}=2(r p_{r}+\frac{\varsigma p_{\theta} \sin \theta}{r^{2}}).
\end{eqnarray}
Now, the second iterative Lagrangian is
\begin{equation}\label{L2}
L^{(2)}=\dot{r}p_r+\dot{\theta}p_{\theta}+\dot{\varphi}p_{\varphi}-\dot{\lambda}_{1} \phi_{1}-\dot{\lambda}_{2} \phi_{2} -H_{c},
\end{equation}
and new symplectic variables and one-form are
  \begin{eqnarray}
  \nonumber {\xi}^{(2)}_{\alpha} &=& (r,\theta,\varphi,p_r,p_{\theta},p_{\varphi},\lambda_{1},\lambda_{2}), \\
  \mathcal A^{(2)}_{\alpha} &=& (p_r,p_{\theta},p_{\varphi},0,0,0,\phi_{1},\phi_{2}),
\end{eqnarray}
with which we can construct the following symplectic two-form,
\begin{eqnarray}\label{f(2)}
f_{\alpha\beta}^{(2)}=
\begin{pmatrix}
   \textbf{0}_{3\times3} & -\textbf{1}_{3\times3} & \textbf{u}_{1\times3}^{T} & \textbf{v}_{1\times3}^{T} \\
   \textbf{1}_{3\times3} & \textbf{0}_{3\times3} & \textbf{0}_{3\times1} & \textbf{w}_{1\times3}^{T} \\
  -\textbf{u}_{1\times3} & \textbf{0}_{1\times3} & 0 & 0 \\
  -\textbf{v}_{1\times3} & -\textbf{w}_{1\times3} & 0 & 0 \\
\end{pmatrix}.
\end{eqnarray}
where, \textbf{v} and \textbf{w} are row matrices which are defined as fallows,
\begin{align}\label{vw}
&  v_{\alpha}=\begin{pmatrix}
 2(p_{r}-\frac{2\varsigma p_{\theta} \sin\theta }{r^{3}})  & \frac{2\varsigma p_{\theta} \cos\theta }{r^{2}} & 0
\end{pmatrix},  \nonumber \\
& w_{\alpha}=\begin{pmatrix}
 2r & \frac{2\varsigma \sin\theta}{r^{2}} & 0
\end{pmatrix}.
\end{align}
The corresponding symplectic two-form is non-singular. Thus, it does not have any null vector and consequently the iterative process stops and no other constraint will be obtained.

Now, we start the symplectic embedding procedure to convert second-class constraints to first ones. The main idea of this procedure is to adjoin Wess-Zumino (WZ) variable to the original phase-space \cite{WZ}. In order to do that, we expand the original Phase-space by introducing a function G as WZ Lagrangian, depending on the original phase-space variables and WZ variable $ \sigma $, as the expansion in terms of WZ variables, defined by
\begin{equation}
G(r,\theta,\varphi,p_r,p_{\theta},p_{\varphi},\lambda_{1},\sigma)=\sum^{\infty}_{n=0}\mathcal{G}^{(n)}.
\end{equation}
This function is gauging potential and satisfies the following boundary condition by vanishing $ \mathcal{G}^{(0)} $,
\begin{equation}\label{G}
G(r,\theta,\varphi,p_r,p_{\theta},p_{\varphi},\lambda_{1},\sigma=0)=0.
\end{equation}

Introducing the new term G into the original symmetrized Lagrangian \eqref{L1}, we obtain a Lagrangian which depends on both original coordinates and WZ variables,
\begin{align}\label{Lt0}
& \tilde{L}^{(1)}=L^{(1)}+L_{WZ}, \nonumber \\
& \qquad=L^{(1)}+G(r,\theta,\varphi,p_r,p_{\theta},p_{\varphi},\lambda_{1},\sigma).
\end{align}
By extending the phase-space, symplectic variables and one-form will be extended as
  \begin{eqnarray}
  \nonumber {\tilde{\xi}^{(1)}}_{\tilde \alpha} &=& (r,\theta,\varphi,p_r,p_{\theta},p_{\varphi},\lambda_{1},\sigma), \\
  \tilde{\mathcal A}^{(1)}_{\tilde \alpha} &=& (p_r,p_{\theta},p_{\varphi},0,0,0,\phi_{1},0).
\end{eqnarray}

Calculating corresponding symplectic two-form we have
\begin{eqnarray}
\tilde{f}_{\tilde\alpha \tilde\beta}^{(1)}= \begin{pmatrix}
f_{\alpha\beta}^{(1)} &  \textbf{0}_{7\times 1}\\
\textbf{0}_{1\times 7} & 0 \\
\end{pmatrix} ,
\end{eqnarray}
which has the following zero modes
\begin{eqnarray}\label{Zeromode1}
\nonumber \tilde{\textit{n}}^{(1)}_{1 \tilde \alpha}=\begin{pmatrix}
\textit{n}^{(1)}_{1 \alpha} & 1
\end{pmatrix}, \\
\tilde{\textit{n}}^{(1)}_{2\tilde \alpha}=\begin{pmatrix}
\textit{n}^{(1)}_{2 \alpha} & 0
\end{pmatrix} .
\end{eqnarray}
These null vectors are the generators of gauge symmetries, since their contraction with the gradient of the potential does not produce any constraint \cite{G2}. One can use the linear combination of these null vectors as,
\begin{equation}\label{Zeromode}
\tilde{\textit{n}}_{\tilde \alpha}=\tilde{\textit{n}}^{(1)}_{1 \tilde \alpha}+\tilde{h}\tilde{\textit{n}}^{(1)}_{2 \tilde \alpha}
\end{equation}

In order to compute $ L_{WZ} $, we must be assured that no other constraint is produced. This mandatory condition generates an iterative system of differential equations, defined by the following equation,
\begin{equation}\label{Gf}
\tilde{\textit{n}}_{\tilde \alpha}\frac{\partial \mathcal{V}^{(1)}}{\partial \tilde{\xi}^{(0)\tilde \alpha}}=\frac{\partial \mathcal{G}^{(n)}}{\partial \sigma}.
\end{equation}
Substituting \eqref{V1} into \eqref{Gf}, we determine $ \mathcal{G}^{(1)} $ after an integration process as
\begin{eqnarray}
\mathcal{G}^{(1)}=(2r p_{r}+\frac{2\varsigma p_{\theta}\sin\theta }{r^{2}})\sigma.
\end{eqnarray}
Putting $ \mathcal{G}^{(1)} $ into \eqref{Lt0}, the first iterative Lagrangian will be obtained.
Hence, for the first-iterated potential we have,
\begin{equation}\label{LV1}
\mathcal{\tilde{V}}^{(1)}=H_{c}-\mathcal{G}^{(1)}.
\end{equation}
Using \eqref{Gf} for the second time to get  $ \mathcal{G}^{(2)} $, we will have,
\begin{eqnarray}
\mathcal{G}^{(2)}=-2(r^{2}+\frac{\varsigma \sin^2\theta }{r^{2}})\sigma^{2}.
\end{eqnarray}
 Substituting $ \mathcal{G}^{(2)} $ into the first iterative Lagrangian, we will obtain the second iterative Lagrangian. Consequently, the second-iterated potential is
\begin{equation}\label{LLV1}
\mathcal{\tilde{V}}^{(1)}=H_{c}-\mathcal{G}^{(1)}-\mathcal{G}^{(2)}.
\end{equation}
Again, using \eqref{Gf} to obtain  $ \mathcal{G}^{(3)} $, we will see that $ \frac{\partial  \mathcal{G}^{(3)}}{\partial \sigma} =0 $, and so, the zero-mode \eqref{Zeromode} does not make a new constraint. In conclusion, all correction terms $ \mathcal{G}^{(n)} $, with $ n\geq 3 $ vanish.
Thus, the gauge invariant canonical Hamiltonian, which had been defined as the symplectic potential, is obtained from
 \begin{align}\label{Hc1}
&\tilde{H}_{(c)}=\mathcal{V}^{(1)}+G(r,\theta,\varphi,p_r,p_{\theta},p_{\varphi},\lambda_{1},\sigma), \nonumber \\
& \qquad=H_{c}+\lambda_{1}\phi_{1}-\mathcal{G}^{(1)}-\mathcal{G}^{(2)},
\end{align}
and the gauged Lagrangian \eqref{Lt0} will be
\begin{equation}\label{Lt}
\tilde{L}^{(1)}=L^{(1)}+\mathcal{G}^{(1)}+\mathcal{G}^{(2)}.
\end{equation}

The generators of infinitesimal gauge transformations can be obtained using $ \varepsilon_{i} \phi_{i} $, where $ \phi_{i}  $ are first-class constraints \cite{Shirzad, Henneaux2}.
Also, substituting zero-modes \eqref{Zeromode1} in the following relation,
\begin{equation}\label{gauge transfor}
\delta\tilde{\xi}^{(1)}_{\tilde \alpha}=\varepsilon_{i} \tilde{n}_{i \tilde\alpha}^{(1)},
\end{equation}
one can obtain the following infinitesimal gauge transformations \cite{G7,kim}.
\begin{align}\label{Gauge transf}
\begin{array}{ll}
  \delta r=0, 	 & \qquad \delta p_{r}=2 r \varepsilon_{1}, \\
  \delta \theta=0, & \qquad \delta p_{\theta}= 2\varsigma \varepsilon_{1} \sin\theta, \\
  \delta \varphi=0, & \qquad \delta p_{\varphi}=0,  \\
  \delta \lambda=\varepsilon_{2}, & \qquad \delta \sigma=\varepsilon_{1},
\end{array}
\end{align}
where, $ \varepsilon_{i} $ are infinitesimal time dependent parameters.Obtaining a nonlinear first-order Lagrangian, we have constructed a gauge theory with the corresponding nonlinear generator functions of gauge transformations for the model. Thus, the gauge symmetry of the model is determined via these transformations. In other words, the gained model is invariant under these transformations.

To obtain gauge symmetries of the model, one can use  the Poisson brackets of the first-class constraints and symplectic variables via the following relation \cite{Shirzad, Henneaux2},
\begin{equation}\label{Shirzad henneaux}
 \delta \tilde{\xi}^{(1)}_{\bar \alpha}=\{\tilde{\xi}^{(1)}_{\bar \alpha},\phi_{j}\}\varepsilon_{j}.
\end{equation}
Apparently, the results obtained from \eqref{Shirzad henneaux} is the same as the infinitesimal gauge transformations \eqref{Gauge transf}.

Considering constrained analysis of the Lagrangian \eqref{Lt} and segregating its corresponding constraints in the following section, we study gauge symmetry of the model more easily.

\subsection{Constraint structure of the gauged Lagrangian}

Using the symplectic method, we enhance the gauge symmetry of the primary model. In following, we derive constraints and phase-space structure of the gauged Lagrangian \eqref{Lt}. In this gauged model, new dynamical variables $\lambda$ and $\sigma$ appear first-orderly in the Lagrangian. So, their momenta are primary constraints in the phase-space. Thus,
\begin{eqnarray}\label{p landa sigma}
& \frac{\partial \tilde{L}^{(0)}}{\partial \dot{\lambda}^{(1)}}=0 : \rightarrow  \rho_{1} = p_{\lambda}, \nonumber \\
& \frac{\partial \tilde{L}^{(0)}}{\partial \dot{\sigma}^{(2)}}=0 : \rightarrow \rho_{2} =p_{\sigma}.
\end{eqnarray}
So, the total Hamiltonian, corresponding to Lagrangian \eqref{Lt}, is
\begin{eqnarray}
\tilde{H}_{T}=\tilde{H}_{c}+\omega_{i}\rho_{i}.
\end{eqnarray}

In the chain-by-chain method \cite{Mojiri}, the consistency of each individual constraint, i.e. $ \rho_{1} $ and $ \rho_{2} $, starts a chain and gives the next element of that chain. Also, the consistency of second-class constraints determines some of Lagrange multipliers, $ \omega^{j} $, while the consistency of first-class ones leads to constraints of the next level,
\begin{eqnarray}
& 0=\{ \rho_{i},\tilde{H}_{T} \}, \nonumber \\
& 0= \{ \rho_{i},\tilde{H}_{c} \}+\omega_{j}\{ \rho_{i},\rho_{j} \}.
\end{eqnarray}
We see that primary constraints are Abelian, i.e. $ \{\rho_{i},\rho_{j} \}=0 $. So, we arrive to secondary constraints $ \psi_{i}=\{ \rho_{i},\tilde{H}_{c}\} $, where $\psi_{1}=\phi_{1}$ and $\psi_{2}=\phi_{2}$.

The consistency of second level of constraints gives no new constraints, Since,
\begin{align}
\begin{array}{ll}
  \{\psi_{1},\tilde H_{c}\}=-\psi_{2}, 	 & \qquad \{\psi_{1},\tilde H_{c} \}\neq 0.
\end{array}
\end{align}
The first relation is identically true on the constrained surface, and the second one determines a Lagrange multiplier due to the fact that $ \{ \psi_{2},\rho_{2} \} \neq 0$.

All in all, we have the following chain structures,
\begin{align}
& \rho_{1} \rightarrow \psi_{1} \rightarrow \psi_{2} \rightarrow \times \quad, \nonumber \\
& \rho_{2} \rightarrow \psi_{2} \rightarrow \times \quad.
\end{align}

Calculating all Poisson brackets, we see that $ \rho_{1} $ is a first-class constraint. The Poisson bracket matrix of other four constraints is non-singular. The non-vanishing elements of that matrix are
\begin{align}
&\{\rho_{2},\psi_{2}\}=\frac{w_{1}^{2}}{4}(4+w_{2}^{2}),\\
&\{\psi_{1},\psi_{2}\}=- \frac{w_{1}^{2}}{4}(4+w_{2}^{2}),
\end{align}
where, $ w_{1} $ and $ w_{2} $ are defined as the components of the row matrix \eqref{vw}.

Since the matrix of Poisson brackets is a square matrix with odd dimensions, it is a singular matrix. This singularity shows that we have more than one first-class constraint in our model, other than $\rho_{1}$. Redefining those constraints and requesting first-class conditions, we will obtain one extra first-class constraint, and in conclusion, there will remain just two second-class constraints.

So, the second first-class constraint is the linear combination of $\rho_{2}$ and $\psi_{1} $, as $\Phi_{3}=\rho_{2}+\psi_{1}$.
This first-class constraint strongly commutes with the two remained second-class constraints, as same as $\rho_{1}$,
\begin{equation}
\{\Phi_{3},\psi_{1}\}=\{\Phi_{3},\psi_{2}\}= 0.
\end{equation}

Therefore, one can rewrite all constraints in the following notation,
\begin{align}\label{phi3}
& \Phi^{(0)}_{1}=p_{\lambda}, \nonumber \\
& \Phi^{(1)}_{1}=\psi_{1}, \nonumber \\
& \Phi^{(1)}_{2}=\psi_{2}, \nonumber \\
& \Phi_{3}=\rho_{2}+\Phi^{(1)}_{1}.
\end{align}
which $ \Phi^{(0)}_{1} $ and $ \Phi_{3} $ are first-class constraints, and $ \Phi^{(1)}_{1} $ and $ \Phi^{(1)}_{2} $ are second-class ones.

Now, we put all second-class constraints into the Hamiltonian to calculate the corresponding Dirac brackets. Also, we take first-class constraints intact, because they obey the Abelian algebra.
 \begin{equation}\label{Hc1f}
\tilde{H}_{c}=H_{c}+\lambda_{1}\Phi^{(1)}_{1}+\sigma \Phi^{(1)}_{2}.
\end{equation}

By counting the dimensions of new variables in extended phase-space, we find that $[\lambda_{1}]=(Length)^{-4}$, and $ [\sigma]=(Length)^{-2}$. Thus, redefining the following variables with length scales,
\begin{equation}
\lambda_{1}=\lambda^{\prime -4},\qquad \qquad \sigma=\sigma^{\prime -2},
\end{equation}
and replacing them in the Hamiltonian \eqref{Hc1f}, we have,
 \begin{equation}\label{Hc'}
\tilde{H}_{c}=H_{c}+\frac{1}{\lambda^{\prime 4}} \Phi^{(1)}_{1}+\frac{1}{\sigma^{\prime 2}} \Phi^{(1)}_{2}.
\end{equation}
We see that two variables with length dimensions have been added to our phase-space. Hence, these length scales extend our configuration space from three to five. As a matter of fact, $ \lambda' $ and $ \sigma' $ can be interpreted as large extra dimensions, which are added to spatial part of the phase-space, via the potential which carries them in the Hamiltonian. This result, i.e. having two extra dimensions, is in a good accordance with \cite{Schmidt}.

\subsection{Quantization of the primary model and the gauged model}

Taking into the account two primary constraints of the original model, $ \phi_{1} $ and $ \phi_{2} $, and two second-class constraints, $ \Phi^{(1)}_{1} $ and $ \Phi^{(1)}_{2} $ of the gauged model, and calculating their corresponding Poisson  brackets matrix, we have,
 \begin{equation} \label{deltaij}
 \Delta_{ij}=\left(
  \begin{array}{cc}
    0 & -\frac{w_{1}^{2}}{4}(4+w_{2}^{2}) \\
    \frac{w_{1}^{2}}{4}(4+w_{2}^{2}) & 0  \\
  \end{array}
\right),
 \end{equation}
 where, $w_{i}$ are the components of the row matrix \eqref{vw}.

In order to determine all Dirac brackets of the original and gauged model, we put the inverse of $ \Delta_{ij} $, in the following formula,
\begin{equation}\label{Dirac brackets}
 \{\xi_{\bar\alpha},\xi_{\bar\beta}\}^{\ast}= \{\xi_{\bar\alpha},\xi_{\bar\beta}\}-\{\xi_{\bar\alpha},\Phi^{(1)}_{i}\} \Delta_{ij}^{-1}\{\Phi^{(1)}_{j},\xi_{\bar\beta}\}.
\end{equation}

Non-vanishing Dirac brackets, using components of the row matrices \eqref{u} and \eqref{vw}, which are common in both primary and gauged models are
\begin{table}[htp]
\begin{center}
\begin{tabular}{|c|c|c|}
  \hline
  Dirac Brackets  &   Primary Model &  Gauged Model \\   \hline
  $(r,p_{r})$ & $1-\frac{4}{4+w_{2}^{2}}$ & $1-\frac{8}{4+w_{2}^{2}}$ \\ \hline
   $(\theta,p_{\theta})$ & $ 1+\frac{ w_{2}^{2}}{4+w_{2}^{2}} $ & $ 1+\frac{2 w_{2}^{2}}{4+w_{2}^{2}} $ \\ \hline
  $(\varphi,p_{\varphi})$ & $ 1 $ & $ 1 $ \\ \hline
  $(r,p_{\theta})$ &  $\frac{4u_{2}}{u_{1}(4+w_{2}^{2})}$ & $\frac{8u_{2}}{u_{1}(4+w_{2}^{2})}$ \\ \hline
  $(\theta,p_{r})$ & $ \frac{4w_{2}}{u_{1}(4+w_{2}^{2})} $ & $ \frac{8w_{2}}{u_{1}(4+w_{2}^{2})} $ \\ \hline
   $ (p_{r},p_{\theta}) $ & $ \frac{4 v_{2}-u_{1}v_{1}w_{2}}{u_{1}(4+w_{2}^{2})} $ & $ \frac{8u_{2}[w_{1}\sigma( 1-4w_{2}^{2})-v_{1}+2\sigma v_{2}]-8 v_{2}}{w_{1}^{2}(4+w_{2}^{2})} $ \\ \hline
  $(\sigma,p_{\sigma})$ & $ N/A $ & $ 1 $ \\ \hline
  $(\lambda,p_{\lambda})$ & $ N/A $ & $ 1 $ \\ \hline
  $(p_{r},p_{\sigma})$ & $ N/A $ & $ 2 u_{1} $ \\ \hline
  $(p_{\theta},p_{\sigma})$ & $ N/A $ & $ -2u_{2} $ \\ \hline
\end{tabular}
\caption{Dirac brackets between extended phase-space variables of the torus model}
\end{center}
\end{table}
\\

As we see, we obtained some non-commutativity in momentum part of phase-space. The momentum-momentum non-commutativity results to  momentum-momentum uncertainty, and in conclusion they lead to one or several minimal momenta, which is a feature of quantization of theories in curved spaces \cite{ml1, Benczik}.

Expanding Poisson brackets of gauged model with respect to the ratio of radii of the torus, and considering $ \varsigma \rightarrow 0 $, the effect of the shape of universe on Poisson structure of phase-space can be studied.
\begin{align}\label{poisson structure}
&  \{ r,p_{r} \}^{\ast} \approx 1+O\left(\varsigma ^2\right), \nonumber \\
&  \{ \theta ,p_{\theta} \}^{\ast} \approx 1+O\left(\varsigma ^2\right), \nonumber \\
&  \{ r,p_{\theta} \}^{\ast} \approx \frac{2 \varsigma  \sin \theta }{r}+O\left(\varsigma ^2\right),\nonumber \\
&  \{ \theta ,p_{r} \}^{\ast} \approx -\frac{2 \varsigma \sin \theta}{r^3}+O\left(\varsigma ^2\right),\nonumber \\
&  \{ p_{r} ,p_{\theta} \}^{\ast} \approx \frac{2 \varsigma  \left(-p_{\theta }\cos \theta -r p_r\sin \theta +4 r^2 \sigma  \sin \theta\right)}{r^3}+O\left(\varsigma ^2\right),  \nonumber \\
&  \{ p_{\theta} ,p_{\sigma} \}^{\ast} \approx -4 \varsigma \sin \theta +O\left(\varsigma ^2\right).
\end{align}
As we see, these Dirac brackets do not have the common canonical structure. 

Also, by characterizing first-class constraints and Dirac brackets of a classical system, its quantized model, say Hilbert space of the quantum states, is fully available at tree level, according to Dirac prescription,
\begin{equation}\label{quant}
\{A,B\}^{\ast} \rightarrow \frac{1}{i \hbar}[A,B], \qquad \hat \phi_{FC}\mid phys> =0,
\end{equation}
where $\hat\phi_{FC}$ is a quantized version of the first-class constraint. Thus, due to \eqref{poisson structure} in quantized model we derive a non-commutative structure in the momentum part of the phase-space. 

\section{Gauging a non-relativistic particle model on the 3-torus}

\subsection{Constraints of a particle on the 3-torus}
As we have mentioned before, a scenario for the universe as a whole is the boundary of a four-dimensional 3-torus. In this section, we consider a test particle on a 3-torus and repeat the previous calculations to derive a gauged model for phenomenological purposes. 

In spherical coordinates, the surface of a torus is defined by
\begin{eqnarray}
  \nonumber x &=& (\check\varsigma_1 \cos\psi+\check\varsigma_2)\cos\varphi, \\
  \nonumber y &=& (\check\varsigma_1\cos\psi+\check\varsigma_2)\sin\varphi, \\
  \nonumber z &=& \check\varsigma_1 \sin\psi \sin \xi, \\
  s &=& \check\varsigma_1 \sin\psi \cos \xi.
\end{eqnarray}
This 3-torus is described by two radii as $\bar \varsigma_{1}$, and $\bar \varsigma_{2}$ \cite{senin,book}. Hence, the primary constraint is described as,
\begin{equation}\label{phi1bar}
\bar\phi_1=r^2 - (\check\varsigma_1 \cos\psi + \check\varsigma_2)^2 - \check\varsigma_1^2 \sin^2\psi,
\end{equation}
where, $ r^2=x^2+y^2+z^2+s^2 $ . This geometrical object best describes the periodic form of a toroidal shape, since it has the simplest geometric form, without any fracture or knot on the surface.

The local properties of this object is recognised by the least possible fundamental lengths, i.e. two radii. From our physical point of view, we get one of these lengths as unit scale, or the Hubble scale $ (\frac{c}{H_0}) $, and the other is measured according to the first.

\subsection{Symplectic analysis of a particle on the 3-Torus}
Here, like the torus model, we have a polynomial which its consistency will determine Lagrange multiplier. Hence, the constrained structure of this model is similar to the previous one, and the only difference occurs in the explicit form of the constraints. Thus, the corresponding gauging process for two models is similar, but because of the greater configuration space for 3-torus, we will obtain a new Poisson structure. Here, we define the corresponding canonical Hamiltonian for unit mass as,
\begin{equation}\label{hambar}
    \bar{H}_{c}=\frac{1}{2} (p_r^2+\frac{p_{\phi
   }^2}{r^2}+\frac{p_{\psi }^2 }{r^2\sin ^2\phi }+\frac{p_{\xi }^2  }{r^2 \sin ^2\phi \sin ^2\psi
   }).
\end{equation}
Due to its configuration space which is affected by a new dynamical variable as an undetermined Lagrange multiplier, $\bar{\lambda}$, which adds a constraint to the free Lagrangian, the corresponding free particle's Lagrangian is singular.
\begin{equation}\label{L0bar}
 \bar{L}^{(0)}=\dot{r}p_r+\dot{\psi}p_\psi+\dot{\phi}p_\phi+\dot{\xi}p_\xi-\bar{H}_c-\bar{\lambda}_{1}\bar{\phi}_{1}(r,\psi,\phi,\xi).
\end{equation}
Symplectic variables and symplectic one-form can be read off from the Lagrangian,
\begin{eqnarray}
  \nonumber \bar{\xi}^{(0)}_{\alpha} &=& (r,\psi,\phi,\xi,p_r,p_\psi,p_\phi,p_\xi,\bar{\lambda}), \\
  \bar{\mathcal A}^{(0)}_{\alpha} &=& (p_r,p_\psi,p_\phi,p_\xi,0,0,0,0,0).
\end{eqnarray}

Starting the symplectic procedure, the corresponding symplectic two-form is obtained as,
\begin{eqnarray}
\bar{f}^{(0)}_{\alpha\beta}=\begin{pmatrix}
 \textbf{0}_{4\times4} & -\textbf{1}_{4\times4} & \textbf{0}_{4\times1} \\
    \textbf{1}_{4\times4} & \textbf{0}_{4\times4} & \textbf{0}_{4\times1} \\
    \textbf{0}_{1\times4} & \textbf{0}_{1\times4} & 0 \\
\end{pmatrix} .
\end{eqnarray}
This matrix is singular and so, it has the following null vector,
\begin{eqnarray}
\bar{n}^{(0)}_{\alpha}=\begin{pmatrix}
 \textbf{0}_{1\times4} &  \textbf{0}_{1\times4} & 1
\end{pmatrix} .
\end{eqnarray}
Using the zero iterative potential,
\begin{eqnarray}
\bar{V}^{(0)}=H_{c}+\bar \lambda_{1} \bar \phi_{1},
\end{eqnarray}
the first constraint will be obtained from following formula.
\begin{equation}\label{consbar}
\bar{\phi}_{1}=\bar{n}_{\alpha}^{(0)}\frac{\partial \bar{V}^{(0)}}{\partial \bar{\xi}^{(0)\alpha}},
\end{equation}
which gives \eqref{phi1bar}.

In order to remove the constraint from the Hamiltonian and add it to the kinetic part of the Lagrangian, we substitute \eqref{consbar} in the Lagrangian \eqref{L0bar}. As a result, the first iterative Lagrangian is,
\begin{equation}\label{L1bar}
\bar L^{(1)}=\dot{r}p_r+\dot{\psi}p_\psi+\dot{\phi}p_\phi+\dot{\xi}p_\xi-\dot{ \bar \lambda}_{1} \bar\phi_{1}- \bar H_{c},
\end{equation}
and the first iterative potential will be,
\begin{equation}\label{Vbar1}
 \bar V^{(1)}= \bar H_{c}.
\end{equation}
Then, new symplectic variables and one-form are
\begin{eqnarray}
  \nonumber { \bar \xi^{(1)}}_{ \alpha} &=& (r,\psi,\phi,\xi,p_r,p_\psi,p_\phi,p_\xi, \bar \lambda), \\
   \bar{\mathcal{A}}^{(1)}_{ \alpha} &=& (p_r,p_\psi,p_\phi,p_\xi,0,0,0,0, \bar \phi_{1}),
\end{eqnarray}
which gives the corresponding symplectic two-form,
\begin{eqnarray}\label{fbar(1)}
 \bar f_{\alpha \beta}^{(1)}=
\begin{pmatrix}
   \textbf{0}_{4\times4} & -\textbf{1}_{4\times4} & \mathcal{U}_{1\times4}^{T} \\
   \textbf{1}_{4\times4} & \textbf{0}_{4\times4} & \textbf{0}_{4\times1} \\
  -\mathcal{U}_{1\times4} & \textbf{0}_{1\times4} & 1 \\
\end{pmatrix},
\end{eqnarray}
and,
\begin{eqnarray}\label{ubar}
 \mathcal{U}_{\mu}=
 \begin{pmatrix}
 2r & 2\check\varsigma_1 \check\varsigma_2 \sin\psi & 0 & 0
\end{pmatrix}
\end{eqnarray}
which $\bar q^{\mu}$ is the spatial component of the symplectic phase-space, i.e. $x, y, z, s$.
Since, the tensor \eqref{fbar(1)} is a singular one, it has following null vectors,
\begin{align}
& \bar{n}^{(1)}_{1 \alpha}=\begin{pmatrix}
\textbf{0}_{1\times 4} & \mathcal{U}_{\mu} & 0
\end{pmatrix},\nonumber  \\
& \bar{n}^{(1)}_{2 \alpha}=\begin{pmatrix}
\textbf{0}_{1\times 4} & \textbf{0}_{1\times4} & 1
\end{pmatrix} .
\end{align}
The linear combination of these null vectors is also a null vector for \eqref{fbar(1)}.
\begin{equation}
\bar{n}_{\alpha}=\bar{n}^{(1)}_{1 \alpha}+\bar h\bar{n}^{(1)}_{2 \alpha}.
\end{equation}
Using \eqref{phi1bar}, we find the second constraint,
\begin{eqnarray}
\bar \phi_{2}=2 (r p_r+\frac{  p_\psi \check{\varsigma}_1 \check{\varsigma}_2 \sin\psi }{r^2 \sin^2\phi }).
\end{eqnarray}
Now, the second iterative Lagrangian will be
\begin{equation}\label{L2bar}
\bar L^{(2)}=\dot{r}p_r+\dot{\psi}p_\psi+\dot{\phi}p_\phi+\dot{\xi}p_\xi-\dot{ \bar \lambda}_{1} \bar \phi_{1}-\dot{\bar \lambda}_{2} \bar \phi_{2} -\bar H_{c} ,
\end{equation}
and new symplectic variables and corresponding one-form are
  \begin{eqnarray}
  \nonumber {\bar \xi^{(2)}}_{\alpha} &=& (r,\psi,\phi,\xi,p_r,p_\psi,p_\phi,p_\xi,\bar \lambda_{1},\bar \lambda_{2}), \\
  \bar{\mathcal{A}}^{(2)}_{\alpha} &=& (p_r,p_\psi,p_\phi,p_\xi,0,0,0,0,\bar \phi_{1},\bar \phi_{2}).
\end{eqnarray}
The corresponding symplectic two-form is
\begin{eqnarray}\label{fbar(2)}
\bar f_{\alpha \beta}^{(2)}=
\begin{pmatrix}
   \textbf{0}_{4\times4} & -\textbf{1}_{4\times4} & \mathcal{U}^{T}_{1\times4} & \mathcal{V}^{T}_{1\times4} \\
   \textbf{1}_{4\times4} & \textbf{0}_{4\times4} & \textbf{0}_{4\times1} & \mathcal{W}^{T}_{1\times4} \\
  -\mathcal{U}_{1\times4} & \textbf{0}_{1\times4} & 0 & 0 \\
  -\mathcal{V}_{1\times4} & -\mathcal{W}_{1\times4} & 0 & 0 \\
\end{pmatrix},
\end{eqnarray}
in which, $\mathcal{V}_{\alpha}$ and $\mathcal{W}_{\alpha}$ are defined as fallow,
\begin{equation}\label{vwbar}
 \mathcal{V}_{\mu}=\frac{\partial \bar \phi_{2}}{\partial \bar q^{\mu}},  \qquad \quad \mathcal{W}_{\mu}=\frac{\partial \bar \phi_{2}}{\partial \bar p^{\mu}}.
\end{equation}
The two-form \eqref{fbar(2)} is non-singular. Thus, it does not have any null vector and consequently there is no other constraint.

To start the symplectic embedding process, we expand the original phase-space, using the unknown function depending on phase-space variables and WZ variable, $ \kappa $, which is defined as the following expansion with the same boundary condition as \eqref{G},
\begin{equation}\label{Gbar}
\bar G(r,\psi,\phi,\xi,p_r,p_\psi,p_\phi,p_\xi,\bar \lambda_{1},\kappa)=\sum^{\infty}_{n=0}  \mathfrak{g}^{(n)},
\end{equation}

Introducing the new term $\bar G $ into the Lagrangian \eqref{L1bar},
\begin{equation}\label{Lt0bar}
\check{L}^{(1)}=\bar L^{(1)}+\bar G(r,\psi,\phi,\xi,p_r,p_\psi,p_\phi,p_\xi,\bar \lambda_{1},\kappa),
\end{equation}
extends the symplectic variables as follows,
  \begin{eqnarray}\label{symplectic var bar 3-torus}
  \nonumber {\check{\xi}^{(1)}}_{\alpha} &=& (r,\psi,\phi,\xi,p_r,p_\psi,p_\phi,p_\xi,\bar \lambda_{1},\kappa), \\
  \check{\mathcal{A}}^{(1)}_{\alpha} &=& (p_r,p_\psi,p_\phi,p_\xi,0,0,0,0,\bar \phi_{1},0).
\end{eqnarray}
Calculating the corresponding two-form symplectic matrix, we have
\begin{eqnarray}
 \check{f}_{\tilde\alpha \tilde\beta}^{(1)}= \begin{pmatrix}
\bar f_{\alpha\beta}^{(1)} &  \textbf{0}_{9\times 1}\\
\textbf{0}_{1\times 9} & 0 \\
\end{pmatrix} ,
\end{eqnarray}
which has the following null vectors,
\begin{eqnarray}\label{Zeromodebar}
\nonumber \check{ {n}}^{(1)}_{1 \tilde\alpha}=\begin{pmatrix}
\bar{n}^{(1)}_{1 \alpha} & 1
\end{pmatrix}, \\
\check{\textit{n}}^{(1)}_{2 \tilde\alpha}=\begin{pmatrix}
\bar{n}^{(1)}_{2 \alpha} & 0
\end{pmatrix} .
\end{eqnarray}
These null vectors also generate gauge transformations on the symplectic variables \eqref{symplectic var bar 3-torus}. To continue the procedure, we use the linear combination of them,
\begin{equation}\label{Zeromodebar1}
\check{\textit{n}}_{\tilde\alpha}=\check{\textit{n}}^{(1)}_{1 \tilde\alpha}+\check{h}\check{\textit{n}}^{(1)}_{2 \tilde\alpha}
\end{equation}
Considering the fact that null vectors \eqref{Zeromodebar} terminates the constraint making process, we can use the following differential equation to obtain $\mathfrak{g}^{(n)}$,
\begin{equation}\label{Gfbar}
\check{n}_{\tilde{\alpha}}\frac{\partial \bar{V}^{(1)}}{\partial \check{\xi}^{(0)\tilde{\alpha}}}=\frac{\partial \mathfrak{g}^{(n)}}{\partial \kappa}.
\end{equation}
Substituting \eqref{V1} into \eqref{Gfbar}, we find $\mathfrak{g}^{(1)} $ as a linear function of WZ variable as,
\begin{eqnarray}
&& \mathfrak{g}^{(1)}=\kappa \bar \phi_{2}\nonumber \\
&& \qquad =2\kappa (p_r r+\frac{\check\varsigma_1 \check\varsigma_2 p_\psi \sin\psi \csc ^2\phi )}{r^2})
\end{eqnarray}
Putting $\mathfrak{g}^{(1)} $ into \eqref{Lt0bar}, the potential becomes
\begin{equation}\label{LV1ba}
\check{V}^{(1)}=\bar H_{c}-\mathfrak{g}^{(1)}.
\end{equation}
Using \eqref{Gfbar} for the second time to get  $ \mathfrak{g}^{(2)} $, we will have
\begin{eqnarray}
&&\mathfrak{g}^{(2)}=-\frac{\kappa^{2}}{2}\{\bar \phi_{1},\bar \phi_{2} \}\nonumber \\
&& \qquad =\frac{-2\kappa^{2}}{r^2} (r^4+\check\varsigma_1^2 \check\varsigma_2^2 \sin^2\psi \csc ^2\phi)
\end{eqnarray}
Substituting $ \mathfrak{g}^{(2)} $ into the first iterative Lagrangian, we obtain the second iterative Lagrangian with the following potential,
\begin{equation}\label{LLV1bar}
\check{V}^{(1)}=\bar H_{c}-\mathfrak{g}^{(1)}-\mathfrak{g}^{(2)}.
\end{equation}

Again, using \eqref{Gfbar} to obtain  $ \mathfrak{g}^{(3)} $, we see that $ \frac{\partial \mathfrak{g}^{(3)}}{\partial \kappa} =0 $ and so, the zero-mode \eqref{Zeromodebar1} does not make a new constraint. Thus, all correction terms $ \mathfrak{g}^{(n)} $ with $ n\geq 3 $ are zero.
So, for the canonical Hamiltonian we have
 \begin{equation}\label{Hc1bar}
\check{H}_{(c)}=\bar H_{c}+\bar \lambda_{1}\bar \phi_{1}-\mathfrak{g}^{(1)}-\mathfrak{g}^{(2)},
\end{equation}
and for the gauged Lagrangian,
\begin{equation}\label{Ltbar}
\check{L}^{(1)}=\bar L^{(1)}+\mathfrak{g}^{(1)}+\mathfrak{g}^{(2)}.
\end{equation}

As we mentioned before, in order to obtain gauge symmetries of the model, one can use  \eqref{Shirzad henneaux} as an option \cite{Shirzad, Henneaux2}, or \eqref{gauge transfor} and corresponding zero-modes \eqref{Zeromodebar} as another one \cite{G7,kim}, which both give the same result.
\begin{align}\label{Gauge transf bar}
\begin{array}{ll}
  \delta r=0, 	 & \qquad \delta p_{r}=\epsilon_{1} 2r, \\
  \delta \psi=0, 	 & \qquad \delta p_{\psi}=\epsilon_{1} 2\check\varsigma_1 \check\varsigma_2 \sin\psi, \\
  \delta \phi=0, 	 & \qquad \delta p_{\phi}=0, \\
  \delta \xi=0, 	 & \qquad \delta p_{\xi}=0, \\
  \delta \bar \lambda=\epsilon_{2},  & \qquad \delta \kappa=\epsilon_{1}.
\end{array}
\end{align}
These are nontrivial variations that make the system invariant under some gauge transformations. So, in the new model, there are some generators for gauge transformations which then we sought.

\subsection{Constraint structure of the gauged Lagrangian of a particle on 3-torus}

Now, we can find the constraint structure of the gauged Lagrangian \eqref{Ltbar}, using the same method as \eqref{p landa sigma} and check consistency conditions. Thus, we have
\begin{eqnarray}
&  \frac{\partial \check{L}^{(0)}}{\partial \dot{\bar \lambda}^{(1)}}=0  : \rightarrow  \check{\rho}_{1} = p_{\bar \lambda}, \nonumber \\
& \frac{\partial \check{L}^{(0)}}{\partial \dot{\kappa}^{(1)}}=0  : \rightarrow \check{\rho}_{2} = p_{\kappa}.
\end{eqnarray}

The constraint structure of the 3-torus is similar to the ordinary torus. Thus, we have the following chain structures.
\begin{eqnarray}
&& \check{\rho}_{1} \rightarrow \check\psi_{1} \rightarrow \check\psi_{2} \rightarrow \times \quad,\nonumber \\
&& \check{\rho}_{2} \rightarrow \check \psi_{2} \rightarrow \times \quad.
\end{eqnarray}

Here, $ \check \rho_{1} $ is a first-class constraint, while its Poisson bracket and all constraints vanish. In order to make another first-order constraint we should redefine them like \eqref{phi3},
\begin{align}
& \check \Phi^{(0)}_{1}=p_{\bar \lambda}, \nonumber \\
& \check \Phi^{(1)}_{1}=\check \psi_{1}, \nonumber \\
& \check \Phi^{(1)}_{2}=\check \psi_{2}, \nonumber \\
& \check \Phi_{3}=\check \rho_{2}+\check \Phi^{(1)}_{1},
\end{align}
which $\check \Phi^{(0)}_{1} $ and $ \check\Phi_{3} $ are first-class, and $ \check \Phi^{(1)}_{1} $ and $ \check \Phi^{(1)}_{2} $ are second-class constraints.

Now, we can make the canonical Hamiltonian,
 \begin{equation}\label{Hc1barf}
\check{H}_{(c)}=\bar H_{c}+\bar \lambda_{1}\check \Phi^{(1)}_{1}+\kappa \check \Phi^{(1)}_{2}.
\end{equation}

Similar to the torus model, the added coordinates to the extended phase-space have the following dimensions,
\begin{align*}
& [\bar \lambda]=(Length)^{-4} , \qquad  [\kappa]=(Length)^{-2} .
\end{align*}
Then, we rewrite the gauged canonical Hamiltonian, using the variables with length dimension as,
 \begin{equation}
\check{H}_{c}=\bar{H}_{c}+\frac{1}{\lambda ^{\prime 4}}\check{\Phi}^{(1)}_{1}+\frac{1}{\kappa^{\prime 2}} \check{\Phi}^{(1)}_{2}.
\end{equation}

As same as torus model, $\lambda'$ and $\kappa'$ can be interpreted as two extra dimensions which are added to the phase-space.

We see that the constrained structure for a free particle on 3-torus is similar to the torus. Thus, its Poisson structure and Dirac brackets are somehow similar. One can obtain non-vanishing Dirac brackets of this model, using \eqref{Dirac brackets}. Also, we can check the effect of 3-torus topology on Dirac brackets by expanding the Poisson brackets of the gauged model with respect to the ratio of radii of the torus, considering both $ \check\varsigma_{1} \rightarrow 0 $ and  $ \check\varsigma_{2} \rightarrow 0 $.

Like torus model, we see that these Dirac brackets do not have the common canonical structure (See Appendix \ref{App B}). Thus, we can interpret such deformations from canonical structure as the effect of topology on the Poisson structure of a particle living on a 3-torus. Also, we can extract some phenomenology from these equations to check non-commutativity structure of the space.

\subsection{Newtonian dynamics of a particle constrained on an asymptotically flat 3-Torus}

As we know, to find Poisson structure between two functions $A$ and $B$ in a phase-space which owns itself a symplectic structure as $ \{ \xi_{\alpha},\xi_{\beta} \}^{\ast}$, one can use the following relation \cite{ml1, ml2}.
\begin{equation}\label{poiss stra}
\{A,B\}=\{ \xi_{\alpha},\xi_{ \beta} \}^{\ast}(\frac{\partial A}{\partial \xi_{\alpha}}\frac{\partial B}{\partial \xi_{\beta}}-\frac{\partial A}{\partial \xi_{\beta}}\frac{\partial B}{\partial \xi_{\alpha}}),
\end{equation}
where $\{ \xi_{\alpha},\xi_{ \beta} \}^{\ast}$ is Dirac bracket between phase-space variables Appendix \ref{App B}.
The relation \eqref{poiss stra} is used to obtain time evolution of phase-space variables and consequently the Hamilton equations of motion.

In order to investigate the effect of topology of the universe on the test particle's dynamics, we rewrite the Hamilton equations of motion for dynamical variables of our gauged model and combine them to obtain the Newton's second law, and more interestingly its corresponding corrections.

To start with, we consider the general radial potential $ V(r)$, added to the Hamiltonian which does not change the derived gauged theory.
\begin{equation}
H=\frac{p_{i} p_{i}}{2}+V(r),
\end{equation}
where $r$ is the radial vector, i.e. $r^2= x^2+y^2+z^2+s^2$, in the Cartesian coordinates.

Hence, due to the nontrivial symplectic structure, the Newton's second law in the direction of $x_{i}$ will be
\begin{equation}\label{accel}
\ddot x_{i}=-\frac{\partial V}{\partial x_{i}}+\mathcal{F}_{i}(\xi_\alpha , \dot \xi_\beta),
\end{equation}
where, $\mathcal{F}_{i}(\xi_\alpha , \dot \xi_\beta)$ can be interpreted as the modification term of the Newton's second law.

Let`s start studying this modification in detail. As we mentioned before, due to the amendments which are imposed by Dirac brackets ascribed to second-class constraints,  equations of motion shall surely include some corrections. Thus, for equations of motion we have,
\begin{align}\label{vel acc}
& \dot x_{i}\approx \{x_{i},H_c\}^*=\{x_{i},\xi_\alpha\}^*\frac{\partial H}{\partial \xi_\alpha}, \nonumber \\
& \ddot  x_{i}\approx \{\{x_{i},H_c\}^*,H_c\}^*.
\end{align}
Calculating the explicit relation of acceleration, i.e. $\ddot x_{i}$, we arrive to
\begin{eqnarray}\label{abc}
&&\hspace*{-0.7cm} \ddot x_{i}\approx \{x_{i},p_{j}\}^* \{ \frac{\partial H_c}{\partial p_{j}},H_c\}^* \nonumber \\
&& +\frac{\partial H_c}{\partial \xi_\alpha}(\{\{x_{i},p_j\}^*\frac{\partial H_c}{\partial p_j},\xi_\alpha\}^*-\{\{\xi_\alpha,H_c\}^*,x_{i}\}^*).
\end{eqnarray}
which is obtained with the the help of Leibniz associative product rule,  
\begin{equation}
 \{F_{1}F_{2},G\}^*=F_{1}\{F_{2},G \}^*+\{F_{1},G \}^* F_{2}.
\end{equation}
All non-vanishing Dirac brackets between extended phase-space variables are obtained in the  Appendix \ref{App B}.
If we switch to the dimensionless coordinates, using following relations,
\begin{eqnarray}
&& r\rightarrow \rho \check\varsigma_1 \qquad , \qquad p_r \rightarrow \frac{p_\rho}{\check\varsigma_1} \nonumber \\
&& \check\varsigma_2 \rightarrow \epsilon \check\varsigma_1 \qquad , \qquad \kappa \rightarrow \hat\kappa \check\varsigma_1^2,
\end{eqnarray}
and expanding the obtained acceleration $ \ddot{r} $ (Appendix \ref{App C}) to the first order of $ \epsilon $, it is easy to investigate that the equations of motion \eqref{abc} in the constructed theory give the following correction for the acceleration of free particle which does not feel any potential. As we know, these corrections are equal to zero for common situations.
\begin{equation}\label{accel0}
\ddot{\rho} \simeq \frac{\epsilon \csc^3 \phi }{\check\varsigma_1^4 \rho^5} (-p_\psi^2 \cos \psi \csc \phi + p_\xi^2 \cot \psi \csc \phi \csc \psi +2p_\phi p_\psi \cos \phi \sin \psi).
\end{equation}
We notice that the acceleration above is independent of the gauge, since to the first order of $ \epsilon $, no $ \hat\kappa $ is included in \eqref{accel0}.

For a particle with unit mass which is affected by the potential in a theory with flat space and the topology $\mathbb{R}^3$, all the terms in \eqref{accel} vanish but the first term $-\frac{\partial V}{\partial x_{i}} $.
So, we obtain the Newton's second law as $\mathcal{F}=\ddot x_{i}$.

On the other hand, for a gauged theory, the relation \eqref{accel0} contains extra terms which have the capability to be explained as the correction of the Newton's second law. 

Also as it is mentioned in Appendix \ref{App B}, if we transform from natural unit to SI by $ X_i=\frac{c}{H_0}x_i $, there is an overall factor $ \frac{H_0}{c} $ multiplied to the correction terms of any new terms, both in equations of motion and symplectic structure.

\subsubsection{A Brief discussion about the Newton's 2nd law}

Here, the reader might be in doubt about the origin of the added terms which are emerged on the RHS of \eqref{abc}. 


As we mentioned before, we modelled our universe in such a way to have a  $ \mathbb{T}^3 $, embedded in $ \mathbb{R}^{4} $, and this is done by specifying a constraint to the surface of 3-torus as $ \bar \phi_{1}(x,y,z,s)=0 $,  which obliges our test particle to live on the flat 4-dimensional surface of 3-torus, not in its 3-dimensional bulk \footnote{ According to \textit{straightening theorem }, one can impose a flat metric on every space \cite{E cal,V del,A es}.}.

So, if we ignore the constraint $ \bar \phi_{1} $  and turn off $\mathbb{T}^3 $, $ \mathbb{R}^4 $ results the euclidean geometry for the universe. In our point of view, the constrained structure of our model covers all necessities to obtain the deviation of trajectory of the test particle in such a universe, and no general relativity calculation is needed any more.

Generally, by starting from Hamilton's equation of motion for the model with the first-classed constraints, and gauged Hamiltonian $ \bar H_c $, one can write the Newton's second law for a particle with unit mass. The velocity of such a particle can be obtained as,
\begin{equation}\label{106}
	\dot{x}_i=\{x_i,\bar H_c+\Lambda_j \phi_j\}^*,
\end{equation}
where, $\Lambda_j$s, as Lagrange undetermined multipliers, are time dependent arbitrary functions. Also, for the time evolution of its momenta we have,
\begin{equation}\label{107}
	\dot{p}_i=\{p_i,\bar H_c+\Lambda_j \phi_j\}^*.
\end{equation}
Here, the RHS of \eqref{106} is defined as $\dot{x}_i=f_i(x_j,p_j,\Lambda_j)$ and we try to find $p_i=f_{i}^{-1}$, where $f_{i}^{-1}$ is the inverted functionality from $p_i$ to $\dot{x}_i$. By defining the RHS of \eqref{107} as $\dot{p}_i=g_i(x_j,p_j,\Lambda_j)$, we have,
\begin{eqnarray}
&&	g_i(x_j,f^{-1}_{j},\Lambda_j)=\frac{d}{dt}f^{-1}_{i} \nonumber \\
&&	\qquad\qquad\qquad =\dot{x}_j\frac{\partial f^{-1}_i}{\partial x_j}+\ddot{x}_j\frac{\partial f^{-1}_i}{\partial \dot x_j}+\dot{\Lambda}_j\frac{\partial f^{-1}_i}{\partial \Lambda_j}.
\end{eqnarray}
Now, we define following matrices as,
\begin{equation}
	\frac{\partial f^{-1}_i}{\partial x_j}=N_{ij}(x,\dot x, \Lambda), \qquad \frac{\partial f^{-1}_i}{\partial \dot x_j}=M_{ij}(x,\dot x, \Lambda),
\end{equation}
to have the equation of motion as, $\textbf{M}\ddot{\textbf{x}}+\textbf{N}\dot{\textbf{x}}+\frac{\partial}{ \partial\Lambda_j} \textbf{f}^{-1}\dot{\Lambda}_j=\textbf{g}$. Here, bold characters notify array and vector factors. Hence, the acceleration is,
\begin{equation}
 \ddot{\textbf{x}}=\textbf{M}^{-1} \textbf{g}-\textbf{M}^{-1}\textbf{N} \dot{\textbf{x}}-\textbf{M}^{-1}\frac{\partial}{ \partial\Lambda_j} \textbf{f}^{-1} \dot{\Lambda}_j.
 \end{equation}
One can have this relation in a simpler form if we go on the constrained surface, which results the acceleration as,
\begin{equation}
 \ddot{\textbf{x}}=\textbf{M}^{-1}\mid_{\phi=0} \textbf{g}-\textbf{M}^{-1}\textbf{N}\mid_{\phi=0} \dot{\textbf{x}}-\textbf{M}^{-1}\frac{\partial}{\partial \Lambda_j} \textbf{f}^{-1}\mid_{\phi=0} \dot{\Lambda}_j.
 \end{equation}
As we see, time-dependent arbitrary functions are remained in the above relation, which shows the deviation of particle's canonical structure, which is the effect of the topology on the dynamics of a test particle.

As a matter of fact, these corrections are classified into two categories. First, terms which are added to \eqref{vel acc} in a proper gauge, to convert weak equalities to strong ones, or in other words, changing $H_c$ to $H_T$, in order to gain full dynamics of the particle. Moreover, for the case of having the common Poisson structure, the only survived factor which is derived from the first term of \eqref{accel}, is $-\frac{\partial V}{\partial x_{i}}$. But, in the model of the particle, living on the torus, as we saw in the last section, we encountered some deviations in the common Poisson structure of phase-space. Thus, aside from the survived first term which adds some corrections itself because of the deformed Poisson structure of the phase-space, there are also correction terms, arose from other terms. All these modification factors can be used to study MOND \cite{Milgram 1,Milgram 2, Milgram 3, Milgram 4},  or to be a candidate to explain dark matter.

In addition, if we consider the gravitational potential of the mass $M$ as $V=-\frac{G M}{r}$, then we can find the correction which is imposed on the universal law of gravity via the attendance of the particle on the torus in tree-dimensional space, and in the presence of the extra gauged coordinates \cite{Kaluza, Arkani, Rubakov}. Thus, one can conclude that for the particle which is affected by the gravitational potential, despite the added corrections for accelerations, other corrections are emerged for the gravitational potential itself.

Moreover, these modification factors can be interpreted as the functions of the new coordinates $ \lambda' $ and $ \kappa' $. From this point of view, we can elucidate these terms as large extra dimensions in Randall-Sundrum scenario \cite{TB3, Randall sundrum2}.

Here, it is worth pointing out that for models dealing with compact extra spatial dimensions, the transition from $ 1/r^2 $ to $ 1/r^4 $ Newtonian gravitation for $ n = 2 $ is calculated \cite{nima1}. This result is comparable to our model since we obtained a model with 2 extra dimensions and a deviation of $ 1/r^4 $ from Newton's inverse square law in \eqref{accel0}. 

\subsection{General relativistic calculations for 3-Torus}
If we make the surface of 3-Torus a subspace $t =constant$ of a 4-dimensional space-time, the metric in coordinate components is defined as follows \cite{senin}.
\begin{eqnarray}
ds^2 = \check{\varsigma}_1^2 [d\psi^2 + (\cos\psi+\frac{\check{\varsigma}_2^2}{\check{\varsigma}_1^2})^2 d\phi^2+\sin^2 \psi d\xi^2].
\end{eqnarray}
Nonvanishing spatial components of Ricci tensor for this model are
\begin{eqnarray}
&& R_{11}=1+\frac{\check{\varsigma}_1 \cos \psi}{\check{\varsigma}_1 \cos \psi +\check{\varsigma}_2},\nonumber \\
&& R_{22}=2 \cos \psi  (\frac{\check{\varsigma}_2}{\check{\varsigma}_1}+\cos \psi),\nonumber \\
&& R_{33} = \frac{\sin ^2\psi  (2 \check{\varsigma}_1 \cos \psi+b)}{\check{\varsigma}_1 \cos \psi+\check{\varsigma}_2}.
\end{eqnarray}
Also, Ricci scalar is calculated as,
\begin{eqnarray}
R = \frac{2 (3 \check{\varsigma}_1 \cos \psi +\check{\varsigma}_2)}{\check{\varsigma}_1^2 (\check{\varsigma}_1 \cos \psi +\check{\varsigma}_2)}.
\end{eqnarray}
This relation will help us to estimate the proper limit which tends the Ricci scalar to zero, and consequently provides the limit for flat space. So, if we get the 3-Torus with large radius $ \check{\varsigma}_1 $ and small radius $ \check{\varsigma}_2 $, one can define,
\begin{eqnarray}
\epsilon = \frac{\check{\varsigma}_2}{\check{\varsigma}_1}\rightarrow 0 \qquad \Rightarrow \qquad R \rightarrow 0.
\end{eqnarray}
Also, one can obtain deviation equations as follows,
\begin{eqnarray}\label{ur}
&& \frac{d^2 r}{d\tau^2}=\sin \psi (-\dot\psi^2 (\frac{\check{\varsigma}_2}{\check{\varsigma}_1}+\cos \psi)+\dot\phi^2 \cos \psi),\nonumber \\
&& \frac{d^2 \psi}{d\tau^2}=\frac{2 \check{\varsigma}_1 \dot r \dot\psi \sin \psi}{\check{\varsigma}_1 \cos\psi+\check{\varsigma}_2} ,\nonumber \\
&& \frac{d^2 \phi}{d\tau^2}= -2 \dot r \dot\phi \cot \psi .
\end{eqnarray}
The components of Einstein tensor for this metric (in coordinate components), without the cosmological constant taken into account, are
\begin{eqnarray}
&& G_{11}=-\frac{\check{\varsigma}_1 \cos \psi }{\check{\varsigma}_1 \cos \psi +\check{\varsigma}_2}, \nonumber \\
&& G_{22}= -\frac{(\check{\varsigma}_1 \cos \psi +\check{\varsigma}_2)^2}{\check{\varsigma}_1^2}, \nonumber \\
&& G_{33}=-\frac{\check{\varsigma}_1 \sin ^2\psi \cos\psi }{\check{\varsigma}_1 \cos \psi+\check{\varsigma}_2}.
\end{eqnarray}

\subsection{Comparing accelerations obtained from constrained systems formalism and general relativity}
Comparing the expanded form of $ \ddot r $ in \eqref{ur} to the first order of $ \epsilon $  as,
\begin{align}\label{ddr}
 \ddot{r}=\frac{\sin \psi  \cos \psi }{r^4}(p_\phi^2 -p_\psi^2 \csc ^4\phi) -\frac{\epsilon \sin \psi }{r^5} \csc ^4\phi (p_\psi^2 r  -2 \check{\varsigma}_1^2 p_r p_\psi \sin \psi  \cos \psi),
\end{align}
and \eqref{accel0}, one can see how these results differ from each other. As a matter of fact, the finite universe which its deviation of acceleration is obtained with general relativistic corrections, includes a term in zeroth order of $ \epsilon $. The advent of this correction term is due to the finiteness of the universe.  On the other hand, for \eqref{accel0}  both lengths, $ \epsilon $ and $ \check{\varsigma_1} $, are included in the acceleration. Therefore, by observing a particle's trajectory one can determine whether the deviation of  acceleration is due to the constraint structure of the model and occurred via gauging procedure, or it is existed due to intrinsically curved manifold.

One should notice that although comparing \eqref{accel0} and \eqref{ddr} shows a substantial difference, they include some terms in 4th order of the inverse of Hubble length, and so measuring their deviation from each other is not an easy task. However, one can find an observable to compare them.

\subsection{Hints for related further researches}

This research can inspire some phenomenological researches for future works. As a good idea, doing a quantum investigation can help us to obtain interesting results. Since, quantum mechanics of the test particle is obtained by converting Dirac brackets in Appendix \ref{App B} to quantum commutator of \eqref{quant}, one can calculate uncertainties between variables in real physical world, and if these uncertainties complied with the quantum commutators extracted from our obtained classical mechanics, we can claim the topology of the universe is the thing which we considered. 

As an example, one could see the following quantum commutator of radial coordinate and corresponding momentum as,
\begin{equation}
[r,p_r] = i \hslash (\frac{\check \varsigma_{1}^2 \check \varsigma_{2}^2 \sin ^2\psi  \csc ^2\phi }{\check \varsigma_{1}^2 \check \varsigma_{2}^2 \sin ^2\psi  \csc ^2\phi+r^4}).
\end{equation}
This form of noncommutativity is the characteristic of quantum mechanics with minimal momentum, which is the generalization of the noncommutative quantum mechanics itself \cite{ml1}.

Also, the obtained quantum mechanics can be investigated by studying the energy spectrum of free particle in this model. As we see, main commutation relations change basically in the gauged particle's quantum mechanics which is affected by a constraint in the configuration space. So, the free particle's Schr\"odinger equation includes an added potential which quantizes corresponding energy spectrum. Therefore, by investigating free particle's energy spectrum and comparing with the very spectrum which is obtained via quantum mechanics of the model, one can determine the corresponding parameters \cite{Benczik,ml2,Akh}.

On the other hand, these forms of commutators and the corresponding quantum mechanics can be investigated in quantum cosmology, especially in inflation model and primordial universe \cite{ijmp,PRD71}. Moreover, highest energy processes which are observed in astrophysics makes us capable to study such a kind of nonommutativity, where the noncommutative parameter affects the dispersion of gamma rays of far astrophysical sources \cite{a44,a46}. Also, canonical noncommutativity deforms infrared photons, which can be investigated via far infrared background radiation (FIBR) investigations \cite{a50,a51,a52,a53}. 

Another way to study such a noncommutativity can be obtained via neutrino observations in the noncommutative sapce. Since, spin $ 1/2 $ particles can have a logarithmic dependence to noncommutative parameter, neutrino kinematics data presents valuable experimental data. On the other hand, since neutrinos are insensitive to dispersion inducing effects, they show spacetime induced effects. Therefore, 1987a-type supernovas with little while (less than a second) and distant (up to $ 10^4 $ Ly) bursts, produce high energy ($\sim$100 MeV) neutrinos which provide testable experimental data \cite{a55}. 

As another result, the advent of the parameter $ \epsilon $ as,
\begin{equation}
\epsilon=\frac{p_{\text{IR}}}{p_{\text{UV}}}=\frac{1/\check \varsigma_{1}}{1/\check \varsigma_{2}}
\end{equation}
is the sign of IR$/$UV mixing, which is the result of noncommutative quantum mechanics. Although its formalism is not straightforward, the field theory based on this particular noncommutative quantum mechanics includes IR$/$UV mixing, which can help us to gain the lower bound of noncommutative scale \cite{a42}. In addition, IR$/$UV mixing obtained via our theory can be investigated in several ways. For instance, it has been shown that IR$/$UV mixing affects CMB spectrum \cite{PRD80}. 

Moreover, in classical point of view and in the case of doing an enough precise measurements, studying the test particle's trajectory can tell us whether the deviation of  acceleration is due to the intrinsic curvature of the universe or the extrinsic one which is induced by the constrained structure of the model. Also, one can obtain the estimate of relative radii of the torus, and in addition one can verify what the topology of the universe would be \cite{Perivolaropoulos}.

Also, since we obtained the deviation from Newton's 2nd law for free particle constrained to the surface of a 3-torus, relation \eqref{accel0},    any test of the gravitational inverse-square law can be used to investigate our model \cite{prl1,prl2}. For instance, to observe the transition from inverse-square law of Newtonian gravitation imposed via extra dimensions in brane world models, some measurements such as submillimeter measurement of gravity are proposed and can be used to test our model \cite{nima1,nima2,Hoyle1,Hoyle2,Bronnikov}.

\section*{Conclusion}

In this article we study the effect of the shape of the universe on a non-relativistic test particle's motion with the help of  formalism of constrained systems,  assuming  that our universe has a topology of a torus. We consider a non-relativistic particle as a test object in this background and find the corresponding gauged phase-space in classical mechanics framework to obtain classical equations of the particle. Also, using Dirac's approach, we gain the particle's quantum mechanics to the first order of $\hbar$ and the associated Hilbert space.

First, we construct a gauge theory on a torus in three-dimensional space as a toy model. We try to concede two degrees of freedom to this particle which lives on a two-dimensional world, using symplectic embedding approach. We show that the interaction which is added to the Hamiltonian of the particle via this process is as the inverse of the length to the power of four on the axis of the gauged degrees of freedom. We also show that these two added degrees of freedom occur due to  imposing the constraint of being on the torus on the Hamiltonian with the help of Lagrange multipliers, and gauging WZ variables. Extracting its Poisson structure, we can quantize the model by replacing Dirac brackets with quantum commutators. Afterwards, we study the effect of the topology on Dirac brackets by enlarging one radius of the torus in comparison with another.

With the help of this toy model, in the main part of this article, we constrain the particle on the hyper-surface of a three-torus in a four dimensional configuration space. Because of the second-class constraints which are imposed on such a particle, the particle will live on a six-dimensional phase-space or three-dimensional configuration space. Then, using the symplectic formalism and embedding phase-space to an extended one, to the particle's  degrees of freedom we add two more ones. These degrees of freedom are equivalent to two second-class constraints, gauged in the final model. By constructing the main brackets of the phase-space, we obtain classical and quantum mechanics of the particle, from which we can interpret deviations occurred from classical and quantum mechanics in $\mathbb{R}^{3}$ space due to the topology of the torus. Hence, we can say that the effect of the topology $ S^{1}\times S^{1}\times S^{1} $ on Dirac brackets, leads to the presence of the particle on a three dimensional manifold, i.e. the particle travels on the surface of the 3-torus and is not floated in its bulk. Thus, obtained quantum mechanics can be investigated by studying the energy spectrum of the free particle in the model. In the gauged particle's quantum mechanics which is affected by the space topology, the main commutation relations change fundamentally. So, even in the free particle's \text{Schr\"odinger} equation, there will be an added potential which makes the corresponding energy spectrum discrete. Hence, by investigating the free particle's energy spectrum and comparing with the very spectrum which is obtained via quantum mechanics of the model, one can determine the corresponding parameters.

Another research work in cosmology can be done with the help of this model to obtain the proposed corrections from modified Newtonian dynamics. Using the equations of motion obtained from the gauged Hamiltonian, we gained the modified form of Newton's second law. We also showed that these correction terms are not only emerged via the intrinsic curvature of the universe, and they also arose from the constrained structure of the model. The obtained correction terms in our model can be compared with the cosmological observations to investigate MOND.

Moreover, the obtained model has two extra gauged degrees of freedom comparing to the common three-dimensional space's degrees of freedom, which add interactions proportion to inverse length square and inverse of the length to the power of four to the Hamiltonian. Particularly, if we consider that the test particle is affected by the gravitational potential of another one in the origin of the space, then in the gauged model, the particle faces the corrections due to the two extra dimensions, which can be related to the brane world cosmology.

\newpage

\appendix
\section{\label{App B}}
Dirac brackets of phase-space variables are shown as follows. By mapping $ r^{2} $ on the constrained surface, i.e. $ r^{2}\mid_{\phi=0} $, one can have these Dirac brackets depending explicitly to the first order of torus radii.

\begin{align*}
&  \{r,p_r\}^*=\frac{\check \varsigma_{1}^2 \check \varsigma_{2}^2 \sin ^2\psi  \csc ^2\phi }{\check \varsigma_{1}^2 \check \varsigma_{2}^2 \sin ^2\psi  \csc ^2\phi+r^4},   \\
&  \{\psi,p_\psi\}^* =\frac{r^4}{\check \varsigma_{1}^2 \check \varsigma_{2}^2 \sin ^2\psi \csc ^2\phi+r^4}, \\
&  \{\phi,p_\phi\}^* =1, \\
&  \{\xi,p_\xi\}^* =1,\\
   &  \{\kappa,p_\kappa\}^* =1, \\
&  \{\lambda,p_\lambda\}^* =1, \\
& \{r, p_\psi\}^*=-\frac{\check \varsigma_{1} \check \varsigma_{2} r^3 \sin \psi}{\check \varsigma_{1}^2\check \varsigma_{2}^2 \sin ^2\psi \csc ^2\phi+r^4},\\
& \{\psi, p_r\}^*=-\frac{\check\varsigma_ 1 \check\varsigma_ 2 r \sin \psi \csc ^2\phi}{\check\varsigma_ 1^2 \check\varsigma_ 2^2 \sin ^2\psi \csc ^2\phi +r^4} ,\\
&  \{p_r,p_\psi\}^* =\frac{\check\varsigma_ 1 \check\varsigma_ 2(\sin\psi 2 \check\varsigma_ 1 \check\varsigma_ 2 \sin \psi \csc ^2\phi  (p_\psi   -2 \check\varsigma_ 1\check\varsigma_ 2 \kappa  \sin\psi)+r^3 (4 r \kappa -p_r))}{\check\varsigma_ 1^2 \check\varsigma_ 2^2 r \sin^2\psi \csc ^2\phi+r^5} \\
& \qquad \qquad +\frac{\check\varsigma_ 1 \check\varsigma_ 2r^2 \cos \psi \csc ^2\phi  (p_\psi -4 \check\varsigma_ 1 \check\varsigma_ 2 \kappa  \sin \psi ))}{\check\varsigma_ 1^2 \check\varsigma_ 2^2 r \sin^2\psi \csc ^2\phi+r^5},\\
&  \{p_r,p_\phi\}^* =\frac{2 \check\varsigma_ 1 \check\varsigma_ 2 r \cot\phi (2 \check\varsigma_ 1 \check\varsigma_ 2 \kappa -p_\psi  \csc\psi )}{\check\varsigma_ 1^2 \check\varsigma_ 2^2+r^4   \csc ^2\psi \sin ^2\phi}, \\
& \{p_\psi, p_\phi\}^* =\frac{2 \check\varsigma_ 1^2 \check\varsigma_ 2^2 \cot \phi  (2 \check\varsigma_ 1 \check\varsigma_ 2 \kappa  \sin \psi -p_\psi )}{\check\varsigma_ 1^2 \check\varsigma_ 2^2+r^4 \csc ^2\psi  \sin ^2\phi}, \\
& \{p_r, p_\kappa\}^*=-2r,\\
& \{p_\psi, p_\kappa\}^* = -2\check\varsigma_ 1 \check\varsigma_ 2 \sin\psi,
\end{align*}

\noindent Now, if we transform from natural unit to standard one in the above Dirac brackets, one can see that the correction terms (cr.t.) added to these brackets are as the order of the inverse of a radius. Here, we get $ [\frac{\text{non-cr.t.}}{\text{cr.t.}}]=[\frac{H_0}{c}] $, therefore,
\begin{equation*}\label{113}
\{x_i, p_{x_{i}}\}^*=\text{non-cr.t.}+\frac{H_0}{c} (\text{cr.t.}).
\end{equation*} 

\newpage
\section{\label{App C}}
With the help of the above Dirac brackets, and relation \eqref{vel acc}, the corresponding accelerations of the test particle including 3-torus radii are,
\begin{align}
&\dot{p}_{r} =\frac {\check\varsigma_ 1 \check\varsigma_ 2 \csc^2\phi} {\check\varsigma_ 1^2 \check\varsigma_ 2^2 r^3 \csc^2\phi + r^7 \csc^2\psi} (\check\varsigma_ 1 \check\varsigma_ 2 p_\phi^2 + 4 \check\varsigma_ 1 \check\varsigma_ 2 r^4 \kappa^2 + p_r p_\psi r^3\csc\psi- 4 p_\psi r^4  \nonumber \\
& \qquad  -  2 p_\phi r^2 \cot\phi (2 \check\varsigma_ 1 \check\varsigma_ 2 \kappa - p_\psi \csc\psi) + \csc^2\phi (-4 \check\varsigma_ 1^2 \check\varsigma_ 2^2 r^2 \kappa^2 \cos\psi+ r^2 \cot\psi \nonumber \\
& \qquad  (4 \check\varsigma_ 1 \check\varsigma_ 2 p_\psi \kappa - p_\psi^2 \csc\psi - p_\xi^2 \csc^3\psi) +\check\varsigma_ 1 \check\varsigma_ 2 (p_\xi^2 \csc^2\psi - (p_\psi - 2 \check\varsigma_ 1 \check\varsigma_ 2 \kappa \sin\psi)^2))  , \nonumber \\
&\dot{p}_{\psi} = \frac{\csc ^2\phi}{\check\varsigma_ 1 ^2 \check\varsigma_ 2 ^2 r^2 \csc^2\phi+r^6 \csc ^2\psi}  (r^3 \cot \psi  (4 \check\varsigma_ 1 ^2 \check\varsigma_2^2 \kappa  (r \kappa -p_r)+\check\varsigma_ 1  \check\varsigma_ 2  p_r p_\psi \csc\psi\nonumber \\
& \qquad  +p_\xi^2 r\csc^4\psi)-2 \check\varsigma_ 1 ^2 \check\varsigma_ 2^2 p_\phi \cot \phi(2 \check\varsigma_ 1  \check\varsigma_ 2 \kappa  \sin\psi-p_\psi)+\check\varsigma_ 1  \check\varsigma_ 2 r (2 \check\varsigma_ 1  \check\varsigma_ 2 (2 \check\varsigma_ 1  \check\varsigma_ 2 \kappa     \nonumber \\
& \qquad  \sin \psi  (r \kappa-p_r)+p_r p_\psi)-p_\xi^2 r \csc^3\psi-p_\psi^2 r \csc\psi))-\check\varsigma_ 1  \check\varsigma_ 2 r^2 \csc\psi (r^2 (p_r \nonumber \\
& \qquad -2 r \kappa )^2+p_\phi^2) , \nonumber \\
&\dot{p}_{\phi}= \frac{\cot\phi \csc ^2\phi }{\check\varsigma_ 1^2 \check\varsigma_ 2^2 r^2 \csc ^2\phi+r^6\csc^2\psi}(\check\varsigma_ 1^2 \check\varsigma_ 2^2 \csc^2\phi
   (p_\xi^2 \csc ^2\psi-(p_\psi-2 \check\varsigma_ 1 \check\varsigma_ 2\kappa \sin\psi)^2)\nonumber \\
& \qquad +r^3 (4 \check\varsigma_ 1^2 \check\varsigma_ 2^2 \kappa    (p_r-r \kappa )-2 \check\varsigma_ 1 \check\varsigma_ 2 p_r p_\psi \csc\psi +p_\xi^2 r \csc ^4\psi +p_\psi^2 r\csc^2\psi)),\nonumber \\
&\dot{p}_{\xi} = 0   
\end{align}

\begin{align}
& \ddot{r}=\frac{1}{(\check\varsigma_ 1^2 \check\varsigma_ 2^2 r \csc ^2\phi+r^5 \csc ^2\psi)^3}\check\varsigma_ 1 \check\varsigma_ 2 \csc^2\phi (\check\varsigma_ 1^4 \check\varsigma_2^4 \csc^6\phi (4 \check\varsigma_ 1^2 \check\varsigma_ 2^2 r^2 \kappa^2 \cos\psi +r^2 \nonumber \\
& \qquad \cot\psi (-4  \check\varsigma_ 1 \check\varsigma_ 2 p_\psi  \kappa +p_\xi^2 \csc^3\psi+p_\psi^2 \csc\psi)+\check\varsigma_ 1 \check\varsigma_ 2 ((p_\psi-2 \check\varsigma_ 1 \check\varsigma_ 2 \kappa  \sin\psi)^2\nonumber \\
& \qquad  -p_\xi^2\csc ^2\psi))+r^4 \csc^2\psi \csc^2\phi(-\check\varsigma_ 1^3 \check\varsigma_ 2^3 (5 p_r^2 r^2+2 p_\phi^2+4 r^4 \kappa ^2)+r^3 \csc\psi \nonumber \\
& \qquad (r \cot\psi (r \csc\psi (4 \check\varsigma_ 1 \check\varsigma_ 2 p_r p_\psi+p_\xi^2 r \csc^3\psi-p_\psi^2 r \csc\psi)  -2 \check\varsigma_ 1^2 \check\varsigma_ 2^2 (p_r^2+2 p_r r \kappa \nonumber \\
& \qquad -2 r^2 \kappa^2))+\check\varsigma_ 1 \check\varsigma_ 2 (4 \check\varsigma_ 1 \check\varsigma_ 2 p_\psi (2    p_r+r \kappa )-p_\xi^2 r \csc ^3\psi-4 p_\psi^2 r \csc\psi))+2 \check\varsigma_ 1^2 \check\varsigma_ 2^2\nonumber \\
& \qquad    p_\phi r \cot \phi (p_\psi r \csc\psi-\check\varsigma_ 1 \check\varsigma_ 2 p_r))+\check\varsigma_ 1^2 \check\varsigma_ 2^2 \csc ^4\phi (r^4 \csc\psi(2 r \cot\psi (2 \check\varsigma_ 1^2 \check\varsigma_ 2^2 \kappa \nonumber \\
& \qquad (2 r \kappa -p_r)+r \csc \psi ) (-2 \check\varsigma_ 1 \check\varsigma_ 2 p_\psi     \kappa +p_\xi^2 \csc ^3\psi +p_\psi^2 \csc (\psi )))+\check\varsigma_ 1 \check\varsigma_ 2 \csc\psi\nonumber \\
& \qquad    (p_\psi^2-2 p_\xi^2 \csc ^2\psi))-\check\varsigma_ 1^3 \check\varsigma_ 2^3 (4 r^3 \kappa  (p_r -r    \kappa )+p_\phi^2))+r^8 \csc ^4\psi (2 p_\phi r \cot\phi\nonumber \\
& \qquad (p_\psi r    \csc\psi-\check\varsigma_ 1 \check\varsigma_ 2 p_r)-\check\varsigma_ 1 \check\varsigma_ 2 (r^2 (p_r -2 r \kappa )^2+p_\phi^2)))
\end{align}


\newpage
\vspace{5cm}
\begin{thebibliography}{99}
\bibitem{PC-12-61} Y.B. Zeldovich, A.A. Starobinsky, Sov. Astron. Lett. 10, 15 (1984).

\bibitem{TA3} D. Stevens, S. Douglas, S. Joseph, Phys. Rev. Lett. 71, 20 (1993).

\bibitem{Cornish0} N.J. Cornish, D.N. Spergel, and G.D. Starkman, Phys. Rev. D 57, 5982 (1998), arXiv:astro-ph/9708225.

\bibitem{Starobinsky} A.A. Starobinsky, JETP Lett. 57, 10 (1993), arXiv:gr-qc/9305019.

\bibitem{Uzan1}  R. Lehoucq, J.P. Uzan, J.P. Luminet, Astron. Astrophys. 363, 1 (2000), arXiv:astro-ph/0005515.

\bibitem{Uzan2} J.P. Uzan, A. Riazuelo, R. Lehoucq, J. Weeks, Phys. Rev. D 69, 043003 (2004).

\bibitem{Riazuelo} A. Riazuelo, J. Weeks, J.P. Uzan, R. Lehoucq, J.P. Luminet, Phys. Rev. D 69, 103518 (2004).

\bibitem{Aurich} R. Aurich, S. Lustig, Mon. Not. Roy. Astron. Soc. 433, 2517 (2013), arXiv:astro-ph/1303.4226.

\bibitem{Cornish1} N.J. Cornish, D.N. Spergel, and G.D. Starkman, Class. Quant. Grav. 15, 2657 (1998).

\bibitem{Cornish2} N.J. Cornish, D.N. Spergel, G.D. Starkman, and E. Komatsu, Phys. Rev. Lett. 92, 201302 (2004), arXiv:astro-ph/0310233.

\bibitem{Cornish3}  J.S. Key, N.J. Cornish, D.N. Spergel, and G.D. Starkman, Phys. Rev. D 75, 084034 (2007), arXiv:astro-ph/0604616.

\bibitem{Cornish4} P.M. Vaudrevange, G. D. Starkman, N.J. Cornish, D.N. Spergel, Phys. Rev. D 86, 083526 (2012), arXiv:1206.2939.

\bibitem{Aurich3} R. Aurich, H.S. Janzer, S. Lustig, F. Steiner, Class. Quant. Grav. 25, 125006 (2008), arXiv:0708.1420.

\bibitem{TA5} M. Tegmark, A. de Oliveira-Costa, A.J.S. Hamilton, Phys. Rev. D 68, 123523 (2003), arXiv:astro-ph/0302496.

\bibitem{Aurich2} R. Aurich, Class. Quant. Grav. 25, 225017 (2008), arXiv:1412.5355.

\bibitem{TD9} J.P. Luminet, M. Lachieze-Rey, Phys. Rep. 254, 135 (1995), arXiv:gr-qc/9605010.

\bibitem{TB1} J. Polchinski, Phys. Rev. Lett. 75, 4724 (1995).

\bibitem{TB2} J. Khoury, B.A. Ovrut, P.J. Steinhardt, N. Turok, Phys. Rev. D 64, 123522 (2001).

\bibitem{TB3} L. Randall and R. Sundrum, Phys. Rev. Lett. 83, 3370 (1999).

\bibitem{TB} R. Murdzek , Int. J. Mod. Phys. D 16, 681 (2007).

\bibitem{TC19} R. Murdzek, Romanian J. Phys., 52, 149 (2006).

\bibitem{Goldestein}  H. Goldstein, C.P. Poole Jr., J.L. Safko, "Classical Mechanics", Pearson Edu. Ltd., (2001).

\bibitem{G1} M. Henneaux, C. Teiltelboim, Quantization of Gauge System, University Press, (1992).

\bibitem{thooft1}  G. 't Hooft, M. Veltman, Nucl. Phys. B 44, 189 (1972).

\bibitem{G2} E.M.C. Abreu, J. Ananias Neto, A.C.R. Mendes, C. Neves, and W. Oliveira, Annalen der Physik 524, 8 (2012), arXiv:1205.7064.

\bibitem{Bergmann} P.G. Bergmann, I. Goldberg, Phys. Rev. 98, 531 (1955).

\bibitem{Dirac1} P.A.M. Dirac, "Lectures on Quantum Mechanics", Belfer graduate School, Yeshiva, Univ. Press, New York, (1964).

\bibitem{Dirac2} P.A.M. Dirac, "Generalized Hamiltonian dynamics", Can. J. Math. 2, (1950).

\bibitem{Monem1} A. Shirzad, M. Monemzadeh,  Phys. Lett.  B 584, 220 (2004), arXiv:hep-th/0311131.

\bibitem{BFFT1} I.A. Batalin and E.S. Fradkin, Nucl. Phys. B 279, 514 (1987).

\bibitem{BFFT2} I.A. Batalin, E.S. Fradkin and T.E. Fradkina, Nucl. Phys. B 314, 158 (1989).

\bibitem{BFFT3} I.A. Batalin and I.V. Tyutin, Int. J. Mod. Phys. A 6, 3255 (1991).

\bibitem{Monem2} A. Shirzad, M. Monemzadeh, Phys. Rev. D 72, 45004 (2005), arXiv:hep-th/0401230.

\bibitem{BFFT4} A.S. Ebrahimi, M. Monemzadeh, Int. J. Theor. Phys. 53, 12 (2014).

\bibitem{FJ1} L. Faddeev, R. Jackiw, Phys. Rev. Lett. 60, 1692 (1988).

\bibitem{FJ2} N.M.J. Woodhouse, "Geometric Quantization", Clarendon Press, Oxford, (1980).

\bibitem{FJ3} J. Ananias Neto, C. Neves and W. Oliveira, Phys. Rev. D 63, 085018 (2001), arXiv:hep-th/0008070v2.

\bibitem{Noether1} M.A. Anacleto, A. Ilha, J.R.S. Nascimento, R.F. Ribeiro and C. Wotzasek, Phys. Lett. B 504, 268 (2001).

\bibitem{Noether2} P.K. Townsend, K. Pilch and P. Van Nieuwenhuizen, Phys. Lett. B 136, 38 (1984).

\bibitem{Noether3} S. Deser, R. Jackiw, Phys. Lett. B 139, 371 (1984).

\bibitem{G4} C. Becchi, A. Rouet and R. Stora, Ann. Phys. [N.Y.] 98, 287 (1976).

\bibitem{G5} I. A. Batalin and G. A. Vilkovisky, Phys. Lett. B 102, 27 (1981).

\bibitem{Monem4} M. Monemzadeh, A.S. Ebrahimi, Mod. Phys. Lett. A 27, 14 (2012).

\bibitem{G7} E.M.C. Abreu, A.C.R. Mendes, C. Neves, W. Oliveira and R.C.N. Silva , JHEP 06, 093 (2013).

\bibitem{Monem3} M. Monemzadeh, A.S. Ebrahimi, S. Sramadi, M. Dehghani, Mod. Phys. Lett. A 29, 1450028 (2014).

\bibitem{G9} J.E. Paschalis, P.I. Porfyriadis, Phys. Lett. B 390, 197 (1997).

\bibitem{kim} Y.W. Kim, C.Y. Lee, S.K. Kim, Eur. Phys. J. C 34, 383 (2004).

\bibitem{Schmidt} C.D. Hoyle, D.J. Kapner, B.R. Heckel, E.G. Adelberger, J.H. Gundlach, U. Schmidt, H.E. Swanson,  Phys. Rev. D70, 042004 (2004), arXiv:hep-ph/0405262.

\bibitem{Mojiri} A. Shirzad, M. Mojiri, Mod. Phys. Lett. A 16, 2439 (2001), arXiv:hep-th/0110023.

\bibitem{WZ} J. Wess, B. Zumino, Phys. Lett. B 37, 95 (1971).

\bibitem{Henneaux} M. Henneaux, C. Teiltelboim, "Quantization of Gauge Systems", Princeton University Press, (1992).

\bibitem{Shirzad} A. Shirzad, M.S. Moghadam, J. Phys. A 32, 46 (1999).

\bibitem{Henneaux2} M. Henneaux, C. Teitelboim, J. Zanelli, Nucl. Phys. B 332, 169 (1990).

\bibitem{ml1} A. Kempf, G. Mangano, R.B. Mann,  Phys. Rev. D52, 1108 (1995), arXiv:hep-th/9412167.

\bibitem{Benczik} S. Benczik, L.N. Chang, D. Minic, N. Okamura, S. Rayyan, T. Takeuchi, Phys. Rev. D66, 026003 (2002).

\bibitem{ml2} S. Benczik, L.N. Chang, D. Minic, N. Okamura, S. Rayyan, T. Takeuchi,  VPI-IPPAP-02-08, (2002), arXiv:hep-th/0209119.

\bibitem{senin} Yu. E. Senin, "Problems of the Theory of Gravitation and Elementary Particles", 13th issue. Moscow: Energoizdat, (1982).

\bibitem{book} J. Plebański, A. Krasiński, "An Introduction to General Relativity and Cosmology", Cambridge University Press, (2006).

\bibitem{E cal} E. Calabi, D. C. Spencer and S. Iyanaga (ed) Papers in Honor of K. Kodaira, (1969).

\bibitem{V del} V. Delecroix, P. Hubert, S. Lelivre, Annales de l’ENS, 47:6 (2014), arXiv:1107.1810.

\bibitem{A es} A. Eskin, J. Chaika, preprint (2013), arXiv:1305.1104.

\bibitem{Milgram 1} M. Milgrom, Astrophys. J. 270, 365 (1983).

\bibitem{Milgram 2} M. Milgrom, Astrophys. J. 698, 1630 (2009).

\bibitem{Milgram 3} M. Milgrom, Phys. Rev. D 89, 024027 (2014).

\bibitem{Milgram 4} M. Milgrom, Can. J. Phys. 93, 107 (2015).

\bibitem{Kaluza} T. Kaluza, Sitzungsber. Preuss. Akad. Wiss. Berlin (Math. Phys.) K1, 966 (1921); O. Klein, Z. Phys. 37, 895 (1926).

\bibitem{Arkani} N. Arkani-Hamed, S. Dimopoulos and G.R. Dvali, Phys. Rev. D59, 086004 (1999), arXiv:hep-ph/9807344.

\bibitem{Rubakov} V.A. Rubakov and M.E. Shaposhnikov, Phys. Lett. B125, 136 (1983).

\bibitem{Randall sundrum2} L. Randall, R. Sundrum, Phys. Rev. Lett. 83, 4690 (1999).

\bibitem{nima1} N. Arkani-Hamed, S. Dimopoulos, G. Dvali, Phys. Lett. B429, 263 (1998).

\bibitem{Akh} R. Akhoury, Y.P. Yao,Phys. Lett. B572, 37 (2003).

\bibitem{ijmp} J. Chluba, J. Hamann, S. P. Patil, Int. J. Mod. Phys. D24, 1530023 (2015).

\bibitem{PRD71} A. Ashoorioon, A. Kempf, R. B. Mann,  Phys. Rev. D71, 023503 (2005), arXiv:astro-ph/0410139.

\bibitem{a40} Sh. Minwalla, M. Van Raamsdonk, N. Seiberg, JHEP 0002 (2000), arXiv:hep-th/9912072.

\bibitem{a44}G. Amelino-Camelia, L. Doplicher, S. Nam, Y.S. Seo, Phys. Rev. D67, 085008 (2003), arXiv:hep-th/0109191.

\bibitem{a46}G. Amelino-Camelia, J. Ellis, N.E. Mavromatos, D.V. Nanopoulos, S. Sarkar, Nature 393, 763 (1998), arXiv:astro-ph/9712103.

\bibitem{a50} R.J. Protheroe, H. Meyer, Phys. Lett. B493 1 (2000).

\bibitem{a51} W. Kluzniak, Astropart. Phys. 11, 117 (1999), arXiv:astro-ph/9905308.

\bibitem{a52} F. Krennrich et al,  Astrophys. J. 560, L45 (2001), arXiv:astro-ph/0107113.

\bibitem{a53} F.A. Aharonian et al, Astron. Astrophys. 393, 89 (2002) , arXiv:astro-ph/0205499.

\bibitem{a55} V. Berezinsky, Phys.Atom.Nucl. 66 (2003) 423-434, Yad.Fiz. 66 (2003), arXiv:astro-ph/0107306.

\bibitem{a42}G. Amelino-Camelia, G. Mandanici, K. Yoshida, JHEP 0401, 037 (2004), arXiv:hep-th/0209254.

\bibitem{PRD80} G.A. Palma, S.P. Patil, Phys. Rev. D 80, 083010 (2009).

\bibitem{Perivolaropoulos}  L. Perivolaropoulos, Phys. Rev. D95, 8, 084050 (2017), arXiv:1611.07293.

\bibitem{prl1} D.J. Kapner, T.S. Cook, E.G. Adelberger, J.H. Gundlach, B.R. Heckel, C.D. Hoyle, H.E. Swanson, Phys. Rev. Lett. 98, 021101 (2007).

\bibitem{prl2} E.G. Adelberger, B.R. Heckel, S. Hoedl, C.D. Hoyle, D.J. Kapner, A. Upadhy, Phys. Rev. Lett. 98, 131104 (2007).

\bibitem{nima2} N. Arkani-Hamed, S. Dimopoulos, G. Dvali, Phys. Rev. D59, 086004  (1999).

\bibitem{Hoyle1} C.D. Hoyle, U. Schmidt, B.R. Heckel, E.G. Adelberger, J.H. Gundlach, D.J. Kapner, H.E. Swanson, Phys. Rev. Lett. 86, 1418 (2001).

\bibitem{Hoyle2} C.D. Hoyle, D.J. Kapner, B.R. Heckel, E.G. Adelberger, J.H. Gundlach, U. Schmidt, H.E. Swanson, Phys. Rev. D70, 042004 (2004).

\bibitem{Bronnikov} K.A. Bronnikov, S.A. Kononogov, V.N. Melnikov, Gen. Rel. Grav. 38, 1215 (2006).


\end{thebibliography}
\end{document}